%% file: ms.tex
\documentclass{aa}

\usepackage{epsfig}
\usepackage{graphicx}
\usepackage{times}
\usepackage{color}
\usepackage{ulem}
\usepackage{hyperref}
\usepackage{url}

\usepackage{xcolor}
\hypersetup{
colorlinks,
linkcolor={red!80!black},
citecolor={blue!80!black},
urlcolor={blue!80!black}
}

\def\cm3{cm$^{-3}$}
\def\kms{km~s$^{-1}$}

\def\lsun{L$_{\odot}$}
\def\rsun{R$_{\odot}$}

\def\msun{M$_{\odot}$}

\def\one{\ts {\,\sc i}}
\def\two{\ts {\,\sc ii}}
\def\three{\ts {\,\sc iii}}

\def\beq{\begin{equation}}
\def\eeq{\end{equation}}

\def\lesssim{\mathrel{\hbox{\rlap{\hbox{\lower4pt\hbox{$\sim$}}}\hbox{$<$}}}}
\def\gtrsim{\mathrel{\hbox{\rlap{\hbox{\lower4pt\hbox{$\sim$}}}\hbox{$>$}}}}

\def\lesssim{\mathrel{\hbox{\rlap{\hbox{\lower4pt\hbox{$\sim$}}}\hbox{$<$}}}}
\def\gtrsim{\mathrel{\hbox{\rlap{\hbox{\lower4pt\hbox{$\sim$}}}\hbox{$>$}}}}

\def\isoni{$^{56}{\rm Ni}$}

\def\one{{\,\sc i}}
\def\two{{\,\sc ii}}
\def\three{{\,\sc iii}}

\def\v1d{{\sc v1d}}
\def\kepler{{\sc kepler}}
\def\mesa{{\sc mesa}}
\def\cmfgen{{\sc cmfgen}}

\def\foe{10$^{51}$\,erg}

\newcommand{\iso}[2]{\ensuremath{^{#1}\rm{#2}}}

\def\aj{AJ}
\def\pasp{PASP}
\def\pasa{PASA}
\def\apj{ApJ}
\def\apjs{ApJS}
\def\apjl{ApJL}
\def\aap{A\&A}
\def\araa{ARA\&A}

\def\mnras{MNRAS}
\def\nat{Nature}

\begin{document}

\title{
Radiative-transfer models for explosions from rotating
and non-rotating single WC stars}

\subtitle{Implications for SN\,1998bw and LGRB/SNe}

\titlerunning{SNe Ic and SNe Ic-BL from single WC stars}

\author{Luc Dessart\inst{\ref{inst1}}
  \and
  D. John Hillier\inst{\ref{inst2}}
  \and
  Sung-Chul Yoon\inst{\ref{inst3}}
  \and
  Roni Waldman\inst{\ref{inst4}}
  \and
  Eli Livne\inst{\ref{inst4}}
  }

\institute{Unidad Mixta Internacional Franco-Chilena de Astronom\'ia (CNRS UMI 3386),
    Departamento de Astronom\'ia, Universidad de Chile,
    Camino El Observatorio 1515, Las Condes, Santiago, Chile.\label{inst1}
    \and
    Department of Physics and Astronomy \& Pittsburgh Particle Physics,
    Astrophysics, and Cosmology Center (PITT PACC),  University of Pittsburgh,
    3941 O'Hara Street, Pittsburgh, PA 15260, USA.\label{inst2}
    \and
    Department of Physics and Astronomy, Seoul National University,
    Gwanak-ro 1, Gwanak-gu, Seoul, 151-742, Republic of Korea.\label{inst3}
    \and
    Racah Institute of Physics, The Hebrew University,
    Jerusalem 91904, Israel.\label{inst4}
  }

  \date{Accepted . Received }

\abstract{
Using 1-D non-Local-Thermodynamic-Equilibrium time-dependent radiative-transfer simulations,
we study the ejecta properties required to match the early and late-time photometric and
spectroscopic properties of supernovae (SNe) associated with long-duration $\gamma$-ray bursts
(LGRBs).
To match the short rise time, narrow light curve peak, and extremely broad spectral
lines of SN\,1998bw requires a model with $\lesssim$\,3\,\msun\ ejecta but a high explosion
energy of a few 10$^{52}$\,erg and 0.5\,\msun\ of \isoni. However the relatively high luminosity,
the presence of narrow spectral lines of intermediate mass elements, and the low ionisation
at the nebular stage are matched with a more standard C-rich Wolf-Rayet (WR) star explosion, with an ejecta
of $\gtrsim$\,10\,\msun, an explosion energy $\gtrsim$\,10$^{51}$\,erg, and only 0.1\,\msun\ of \isoni.
As the two models are mutually exclusive, the breaking of spherical symmetry is essential to match
the early/late photometric/spectroscopic properties of SN\,1998bw.
This conclusion confirms the notion that the ejecta of SN\,1998bw is aspherical on large scales.
More generally, with asphericity, the energetics and \iso{56}Ni mass of LGRB/SNe are reduced
and their ejecta mass is increased, favoring a massive fast-rotating Wolf-Rayet star progenitor.
Contrary to persisting claims in favor of the proto-magnetar model for LGRB/SNe,
such progenitor/ejecta properties are compatible with collapsar formation.
Ejecta properties of LGRB/SNe inferred from 1D radiative-transfer modeling
are fundamentally flawed.
}


\maketitle
\label{firstpage}

\section{Introduction}
\label{sect_intro}

Although some uncertainty remains on the properties of Type Ibc supernovae (SNe) and their progenitors,
the general consensus
is that they arise from H-deficient  stars with Wolf-Rayet (WR) like abundances, and with a final mass
at core collapse of $\sim$\,5\,\msun. Their ejecta have a kinetic energy of $\sim$\,10$^{51}$\,erg, quite
typical of what is inferred
for H-rich core-collapse SNe \citep{ensman_woosley_88,drout_11_ibc,taddia_ibc_15,prentice_ibc_16}.
Due to their low mass most of them are thought to arise in binary systems in which one, or both stars,
have undergone mass transfer \citep{podsiadlowski_92,eldridge_08_bin}.
The fate of single WR stars, which have a large final mass at collapse \citep{pac_wr_07},
is however unclear \citep{D15_WR}.
Type Ib and Ic SNe are distinguished by the presence/absence
of He\one\ lines in optical spectra, perhaps because Type Ic progenitors
have a smaller mass of He and a larger CO core (see, for example,
\citealt{georgy_snibc_09}, \citealt{dessart_11_wr}, \citealt{liu_snibc_15}).

The broad-line (BL) Type Ic SNe (hereafter SNe Ic-BL)
are observationally distinct from normal Type Ibc SNe.
Their spectral lines are so broad at early times that line identifications are compromised, and the
velocity at maximum absorption is uncertain. Nonetheless, their optical spectra exhibit 3 to 4
smooth ripples that imply a large expansion rate with a photospheric velocity that may approach 30,000\,\kms\
prior to maximum (for a recent discussion, see \citealt{modjaz_blic_15}).
Their large expansion rate,  their association in some cases with long-duration $\gamma$-ray bursts
(LGRB),\footnote{All LGRB/SNe are broad lined Type Ic SNe but not all SNe Ic-BL are
associated with an LGRB.}
and their large luminosities, suggest that a subset of core-collapse SNe produce much larger explosion energies
and a much greater mass of \iso{56}Ni (see, e.g., \citealt{iwamoto_98bw_98}; \citealt{woosley_98bw_99}).

One notorious example presenting these extraordinary properties is SN\,1998bw,
which is an LGRB/SN \citep{galama_98bw_98}.
Using 1-D radiation hydrodynamics simulations and OPAL opacities, \citet{woosley_98bw_99} find that
a $\sim$4.77\,\msun\ CO-rich ejecta with 0.49\,\msun\ of \isoni\ and a kinetic energy of
2.8$\times$\,10$^{52}$\,erg yields a satisfactory match to the light curve of SN\,1998bw prior to $\sim$\,100\,d.
This model corresponds to a 6.55\,\msun\ C-rich  WR progenitor star leaving behind a 1.78\,\msun\  neutron star remnant.
They find that higher mass models yield a too long rise time.
Using a more simplified light curve model with a prescribed opacity, \citet{iwamoto_98bw_98}
argue for a $\gtrsim$\,10\,\msun\ ejecta with 2--5$\times$\,10$^{52}$\,erg. This is more than twice
the ejecta mass of  \citet{woosley_98bw_99}.
An alternative scenario invokes an asymmetric explosion that produces ejecta with an energy
of 2$\times$\foe, a total mass of 2\,\msun, and a \iso{56}Ni mass of 0.2\,\msun\ \citep{hoeflich_98bw_99}.
However, all these models underestimate the luminosity at late times because
the ejecta density is too low to sufficiently trap $\gamma$-rays from \iso{56}Co decay
\citep{sollerman_98bw_00,patat_98bw_01}.
A similar discrepancy was found for SN\,1997ef, which led \citet{mazzali_97ef_00} to propose
that the explosion may have been asymmetric.

\citet{maeda_98bw_03} emphasise the inherent problem of spherically-symmetric models to explain
simultaneously the early-time and late-time light curve of SN\,1998bw and other LGRB/SNe.
They propose a two component model with \iso{56}Ni both located in the outer fast-moving ejecta
and in the inner denser and slower ejecta. With this additional freedom, the model can explain the huge
brightness at early times and the sustained brightness at late times.
A strong ejecta asphericity is also inferred by \citet{maeda_neb_06,maeda_neb_08} from nebular
phase spectra of SN\,1998bw and other LGRB/SNe.

\input{tab_prog_prop}

\input{tab_sn_prop}

Fast rotation is thought to be central for the production of LGRB/SNe, both in the
context of the collapsar model \citep{woosley_93} and the proto-magnetar model
\citep{wheeler+00,bucciantini+08,metzger+11}.
To preserve its original angular momentum, an initially fast rotating single star must lose little mass.
Furthermore, because of its redistribution within the star, angular momentum
may be retained if  the star does not evolve through a giant phase, which occurs if it evolves chemically
homogeneously (this holds even in the absence of mass loss;
\citealt{yoon_grb_05}; \citealt{WH06}; \citealt{yoon_grb_06}).
With the current understanding of hot star winds, this situation requires an evolution
at low metallicity. An important corollary is that then the final mass is close to the initial mass.
As discussed above, because the inferred ejecta mass of LGRB/SNe may be only
a few \msun, this poses serious concerns about the whole theory.
% An alternative scenario is to assume LGRB/SNe arise in binary systems \citep{cantiello_grb_07},
% or possibly from a merger \citep{pod_ce_10}.

With the goal of reaching a final mass at collapse of about 10\,\msun\ and keeping a lot
of angular momentum, \citet{WH06} evolved 12-16\,\msun\ stars at very low metallicity.
These stars die with lots of angular momentum but their final mass is low and their
propensity for black-hole (BH) formation is questionable. Indeed, \citet{D12_BH} argue
that nearly all of these models are unlikely to form a BH, primarily because their
core structure (or compactness) is analogous to that of standard 15\,\msun\ red-supergiant (RSG)
stars that are expected to produce SNe II-P.
In the study of \citet{WH06}, only the more massive models  (models 35OB-35OC), with a progenitor
mass of 35\,\msun\ have a huge compactness and
may produce a BH (if they do not explode through a magneto-rotational explosion
first; \citealt{D08_GRB}).
However, models 35OB-35OC die with a final mass of 20-30\,\msun, which is
incompatible with the inferences of \citet{woosley_98bw_99} or \citet{iwamoto_98bw_98}
for  SN\,1998bw.
The model of  \citet{woosley_98bw_99} leaves behind a neutron star, or probably a magnetar,
but not a BH. \citet{yoon_grb_06} studied the properties of
low-metallicity fast rotating massive stars. These stellar models evolve chemically homogeneously,
retain a large amount of angular momentum, but produce WR stars at death with a mass of 10--30\,\msun.

This issue with the final WR mass is major. It is a fundamental characteristic of the progenitor; it controls
the ejecta properties and is likely determinant for the nature of the compact remnant.
If one could invoke a large final WR mass for LGRB/SN
progenitors, numerous concerns would immediately disappear for the production of a collapsar
\citep{woosley_93,MacFadyen_collapsar_99}.
First, such models are more likely to form BHs. As they no longer need to lose much
mass they may retain a lot of angular momentum, although a low metallicity would still be needed to limit
the wind mass-loss rate and to ensure a chemically-homogeneous  evolution (to prevent a giant phase).
Massive WRs are naturally encountered in the high-luminosity compact regions,
which are the site of LGRB/SNe
(see, e.g., \citealt{lefloch_grb_03}, \citealt{fruchter_grb_06}, \citealt{modjaz_grb_08}).

Because of the complexity of LGRB/SNe, it is important to perform independent studies
of their light curve and spectra with different tools and techniques.
Even today, the bulk of the analyses of LGRB/SNe and standard
SNe Ic is based on a simplistic 1D modelling of the LC and/or no modelling of the
spectra (for recent studies, see, e.g.,
\citealt{drout_11_ibc}; \citealt{dhelia_13dx_15}; \citealt{toy_13dx_16};
\citealt{prentice_ibc_16}; \citealt{volnova_13dx_17}).
There is a persisting controversy with the origin of LGRB/SNe. Some recent
studies argue for the proto-magnetar model \citep{mazzali_pm_14, wang_pm_17}
and it is therefore necessary to independently assess LGRB/SN ejecta properties and
address the pros and cons of the collapsar and the proto-magnetar models.

Here, we present 1-D time-dependent radiative transfer models based on the explosion
of  a set of carbon-rich WR stars, which result from the evolution of a rotating or
non-rotating 40\msun\ star on the zero-age-main-sequence (ZAMS). Unlike previous studies, our approach computes
the evolution of the full ejecta from early to late times, and generates  the
frequency-dependent emergent flux, from which both multi-band light curves and UV/optical/near-IR
spectra can be simultaneously extracted --- the  same physics is used to compute both photometric and spectroscopic observables \citep{HD12}.
This is the first time such simulations are presented.

The paper is organised as follows.
In the next section, we present the numerical approach, including the simulation of the pre-SN
evolution, the explosion dynamics, and the radiative transfer.
These simulations are similar to the CO core explosion models of \citet{woosley_98bw_99} --- with
the exception of our BH-forming SN model ---  but the observables are computed with greater
physical consistency.
In Section~\ref{sect_res}, we discuss the key photometric and spectroscopic properties of these models
and confront them to the observations of SN\,1998bw.
We confirm that the early-time LC requires a large ejecta energy-to-mass ratio and a large \iso{56}Ni mass,
and that such a model fails to match the late time LC.
Further, our models that match the early-time spectra fail to match even approximately
the late-time spectra, and vice-versa.
In support of previous studies, we find numerous lines of evidence that the SN\,1998bw may be
axisymmetric, both from the spectra and the light curve, and both from early times and late times.
In Section~\ref{sect_conc}, we present our conclusion.
This study is a stepping stone for forthcoming radiative transfer simulations in 2-D.

\begin{figure}
\epsfig{file=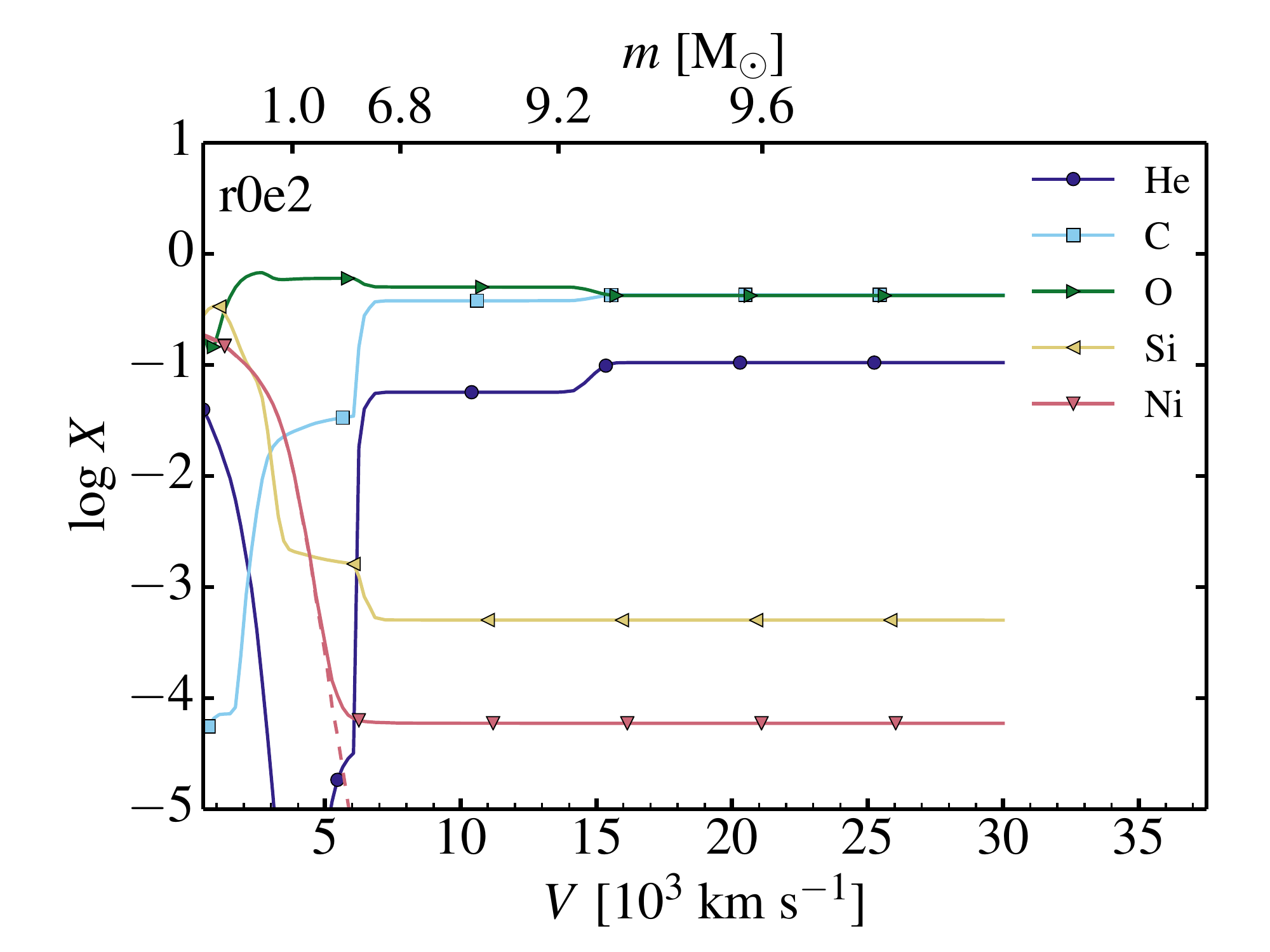,width=8cm}
\epsfig{file=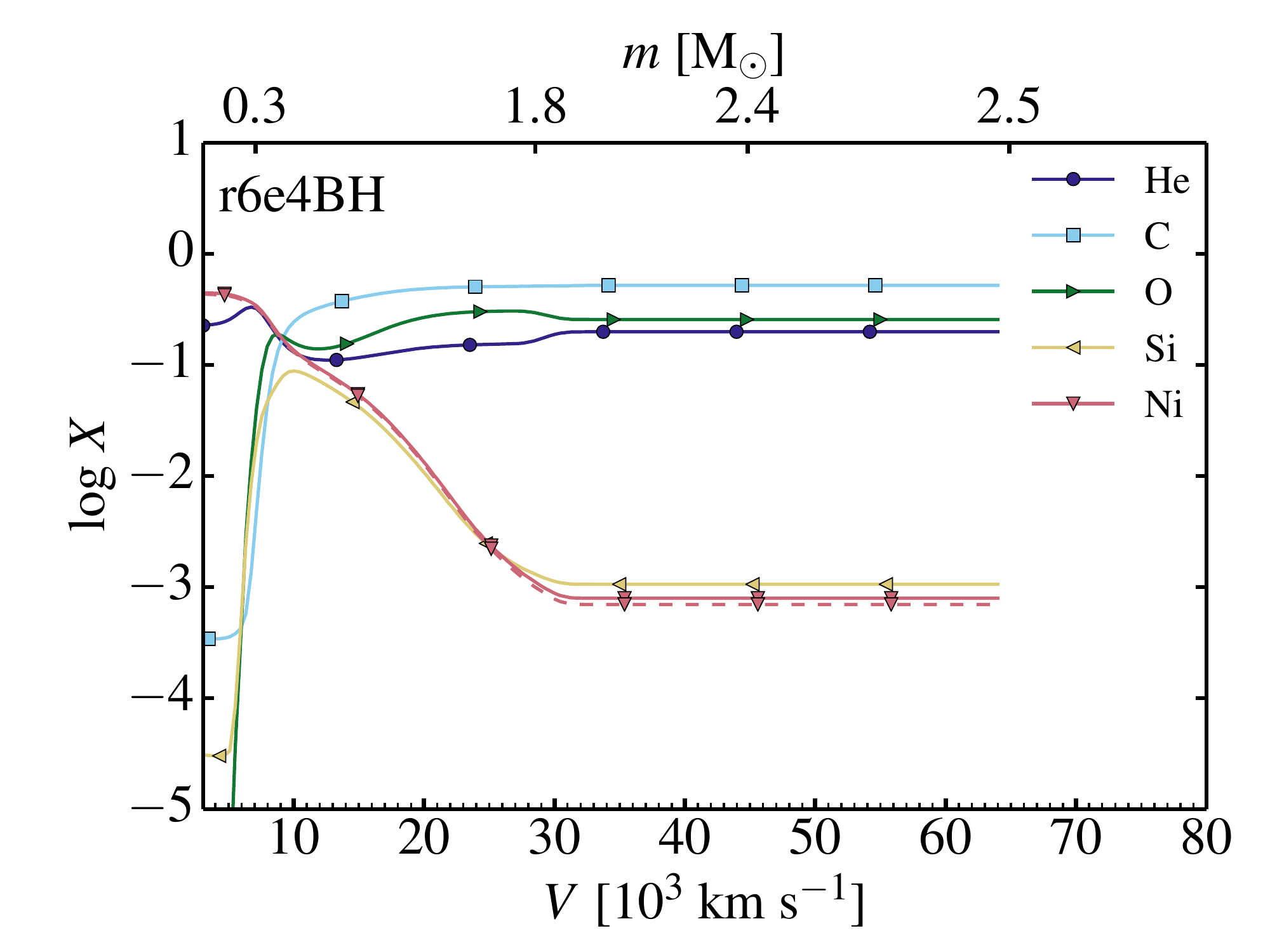,width=8cm}
\caption{
Ejecta composition at the start of the \cmfgen\ simulations for
models r0e2 (top; $M_{\rm e}=$\,9.69\,\msun; post-explosion time of $\sim$\,1.1\,d)
and r6e4BH (bottom; $M_{\rm e}=$\,2.55\,\msun; post-explosion time of $\sim$\,4.6\,d).
The dashed line corresponds to the \iso{56}Ni mass fraction at the given time. The top axis
gives the Lagrangian mass coordinate in the ejecta (zero corresponds to the base of the ejecta).
In model r6e4BH, most of the O-rich shell  has been accreted into the 4\,\msun\ BH.
In particular, the outer layers where He is present now overlap with the \iso{56}Ni-rich regions.
\label{fig_init_comp}
}
\end{figure}

\section{Numerical Setup}
\label{sect_setup}

The simulations we present in this study are all based on a 40\,\msun\ progenitor star evolved
from the ZAMS until the onset of core collapse with the code
\mesa\ \citep{mesa1,mesa2,mesa3}, version 5696.
The metallicity is set to $0.0162$ for the solar metallicity models, and 0.002
for the low metallicity model. We consider models with and without rotation on the
ZAMS. The effects of rotation
on stellar structure, chemical mixing, angular momentum transport, and mass loss
are accounted for according to the methods available in \mesa\ \citep{mesa3}.
All models correspond to single stars, so that any mass loss takes the form of a radiation-driven wind.
The wind mass loss rate is determined by the `Dutch' recipe in \mesa,
combining the values from  \citet{glebbeek+09}, \citet{Nieuwenhuijzen90},
\citet{nugis_lamers_00}, and \citet{vink_mdot_01},
with a coefficient $\eta=1.4$.  Convection is followed according to the Ledoux criterion,
with a mixing length parameter $\alpha_{\rm MLT}=$\,3, a semi-convection efficiency parameter
$\alpha_{\rm sc}=$\,0.1 \citep[Eq.~12]{mesa3}, and an exponential overshoot
with parameter $f=$\,0.008 \citep[Eq.~2]{mesa1}.

Since all models derive from a 40\,\msun\ star on the ZAMS, our nomenclature
distinguishes models according to the initial rotation rate and the metallicity.
We use r0 to refer to a model without rotation, and r4 (r6) for a model
spinning initially at 40\% (60\%) of the critical angular velocity
$\Omega_{\rm crit}$ at the surface.
We also include one fast-rotating model at a tenth of the solar metallicity (model r6z).
By the time of core collapse, all models are H free, but they retain a small amount of He
within the CO core. They would correspond to C-rich (or O-rich) WR stars
\citep{maeder_wr_94,langer_wr_94,groh_13_presn}.
Some model properties at the onset of core collapse are given in Table~\ref{tab_prog_prop}.

Using \v1d\ \citep{livne_93,DLW10a,DLW10b}, the explosion is triggered by means of a piston
to produce an ejecta kinetic energy of about 4 and 12$\times$\,10$^{51}$\,erg
(models are identified by suffixes e2 and e4 in order of explosion energy).
The code treats nuclear burning but the explosive nucleosynthesis yields a
\iso{56}Ni mass that depends sensitively on the piston location (in mass space) and on
the piston trajectory/velocity. Consequently, our simulations produce a sizeable scatter
in \iso{56}Ni mass (not necessarily strongly correlated with explosion energy), with values
in the range 0.085--0.696\,\msun.
In all models except r6e4BH, the piston mass cut is located at the edge of the Fe core, somewhere between
1.5--2.0\,\msun\ (producing a neutron star remnant).
In model r6e4BH, we place it at 4\,\msun\ to produce a BH and explore the effect of
a much lower ejecta mass (i.e., 2.55\,\msun, compared to 5-10\,\msun\ for the rest of the sample)
on the SN radiation. This could occur in a progenitor star whose innermost stable
circular orbit is located at 4\,\msun.
The sensitivity of our results to BH mass and for a set of fast-rotating carbon-rich WR progenitors
will be presented in a forthcoming paper.

\begin{figure}
\epsfig{file=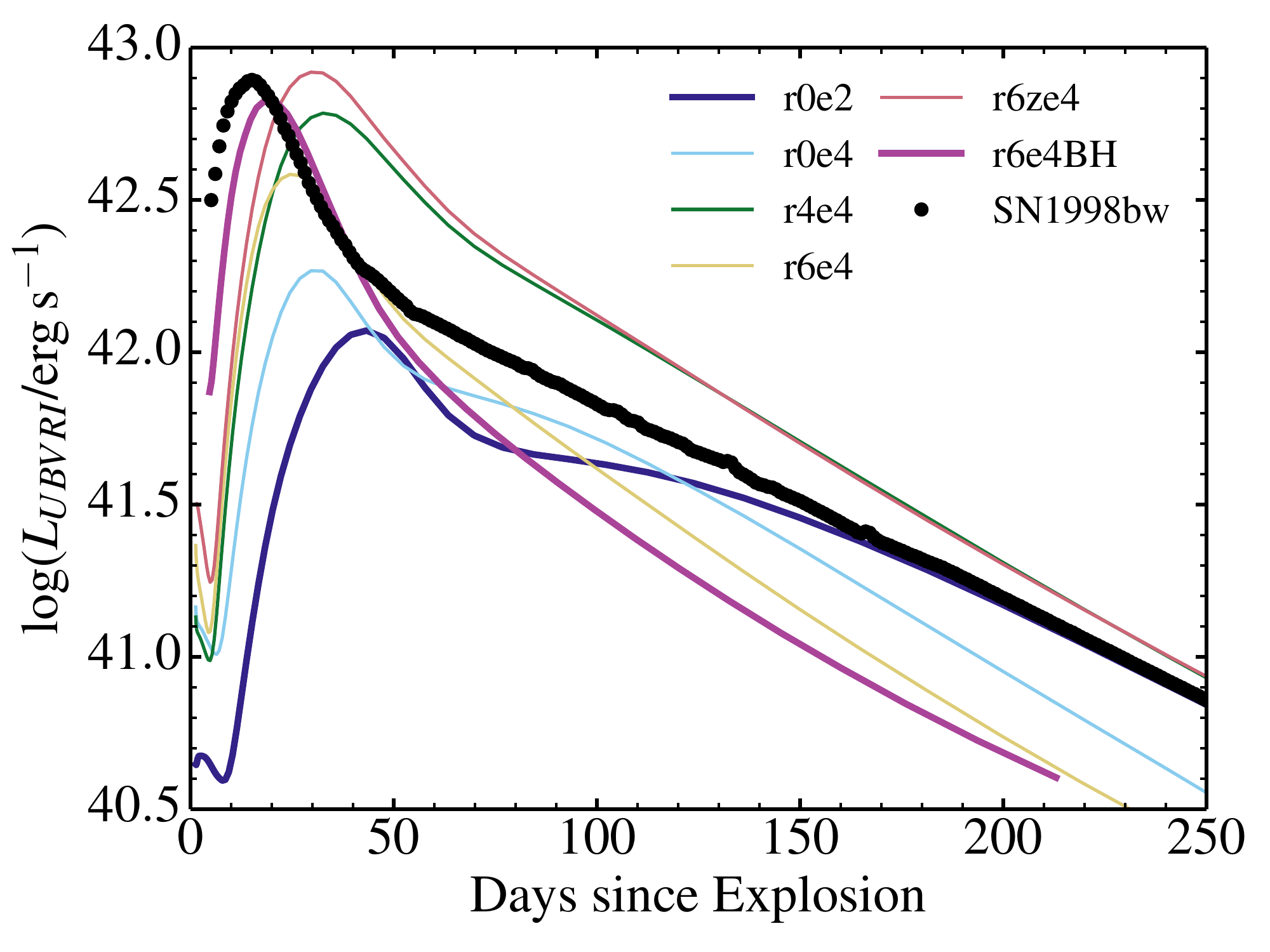,width=8.5cm}
\caption{Evolution of $L_{\rm UBVRI}$ for SN\,1998bw and for the set of single WR star explosion
models  (with and without rotation -- see Tables~\ref{tab_prog_prop} and \ref{tab_sn_prop}
for details).
Models r4e4, r6e4, r6ze4, and r6e4BH come close to match the luminosity around maximum
(with an  offset in rise time), but underestimate the luminosity at late times.
Less energetic and more massive models with a lower  \iso{56}Ni mass (models r0e2 and r0e4)
are in better agreement at late times, but they fail to match the phase around maximum
(see Section~\ref{sect_res} for discussion).
\label{fig_lum_set_1}
}
\end{figure}

In models forming a neutron star, the piston is given a fixed outward velocity
of  15000-30000\,\kms\ until the total energy of the stellar envelope equals
the desired asymptotic kinetic energy (the piston is set to rest thereafter).
In model r6e4BH, we instead adopt the same piston trajectory as in \kepler\ \citep{ww95}.
Placing the piston at $M_{\rm piston}=$\,4\,\msun, we first drive the piston inwards for 30\,s to simulate
the envelope collapse into the gravitational potential well. When the piston reaches down to
$R_{\rm min}=$\,500\,km, we drive the piston outwards at an initial speed $V_{\rm 0}$ of 40000\,\kms.
The piston trajectory follows
$dR_{\rm piston}/dt =
\left[ \alpha G  M_{\rm piston}(1/R_{\rm piston} - 1/R_{\rm min})+  V_{\rm 0}^2  \right]^{1/2}$,
where $\alpha = 0.4 V_{\rm 0}^2 / M_{\rm piston}$ (see \citealt{ww95} for discussion). The collapse phase
brings the inner envelope to small radii, compressing the material to large densities. When the piston moves
outwards, the large post-shock temperatures necessary for the synthesis of a large mass of \iso{56}Ni
are easily produced.

\begin{figure*}
\epsfig{file=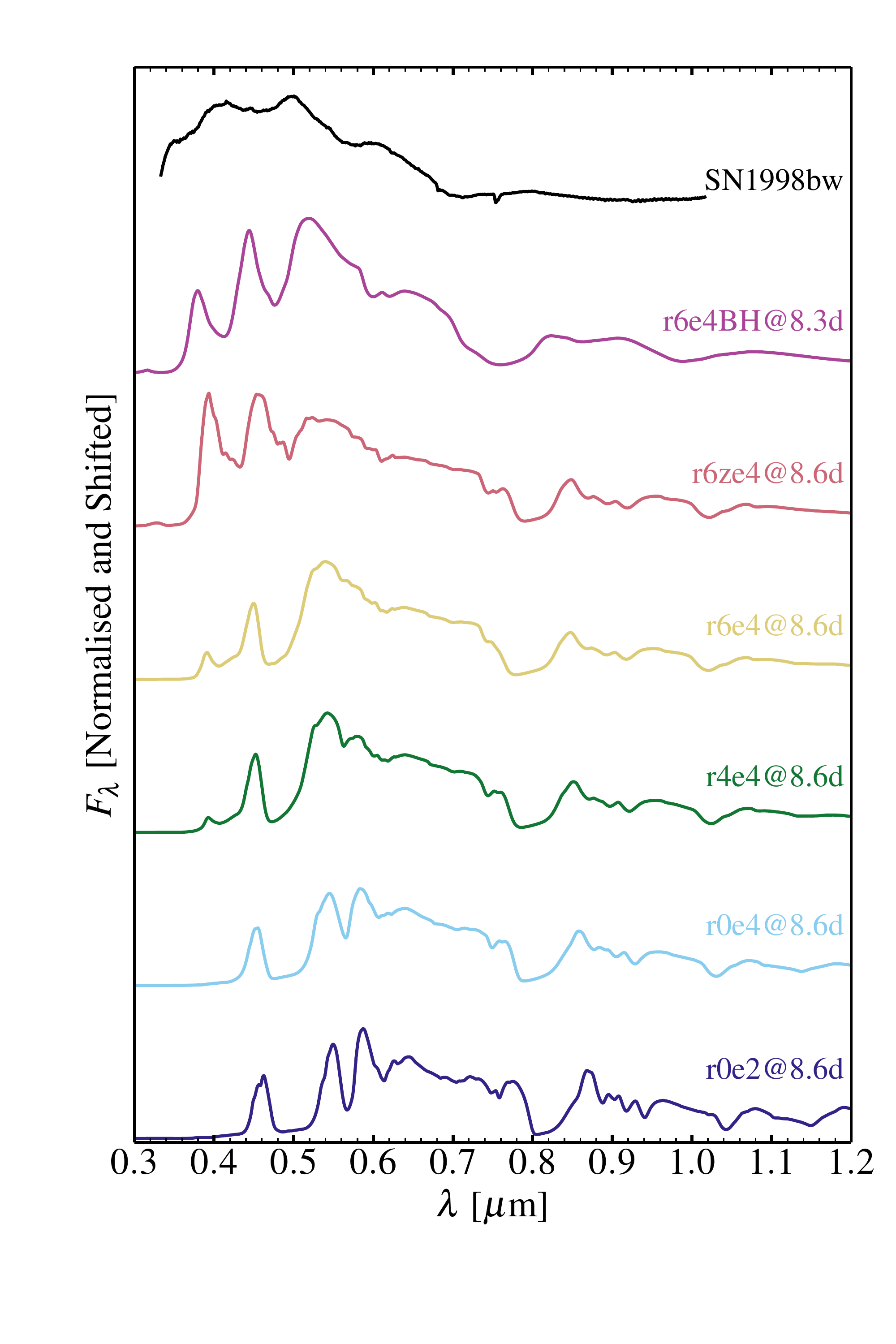,width=8.5cm}
\epsfig{file=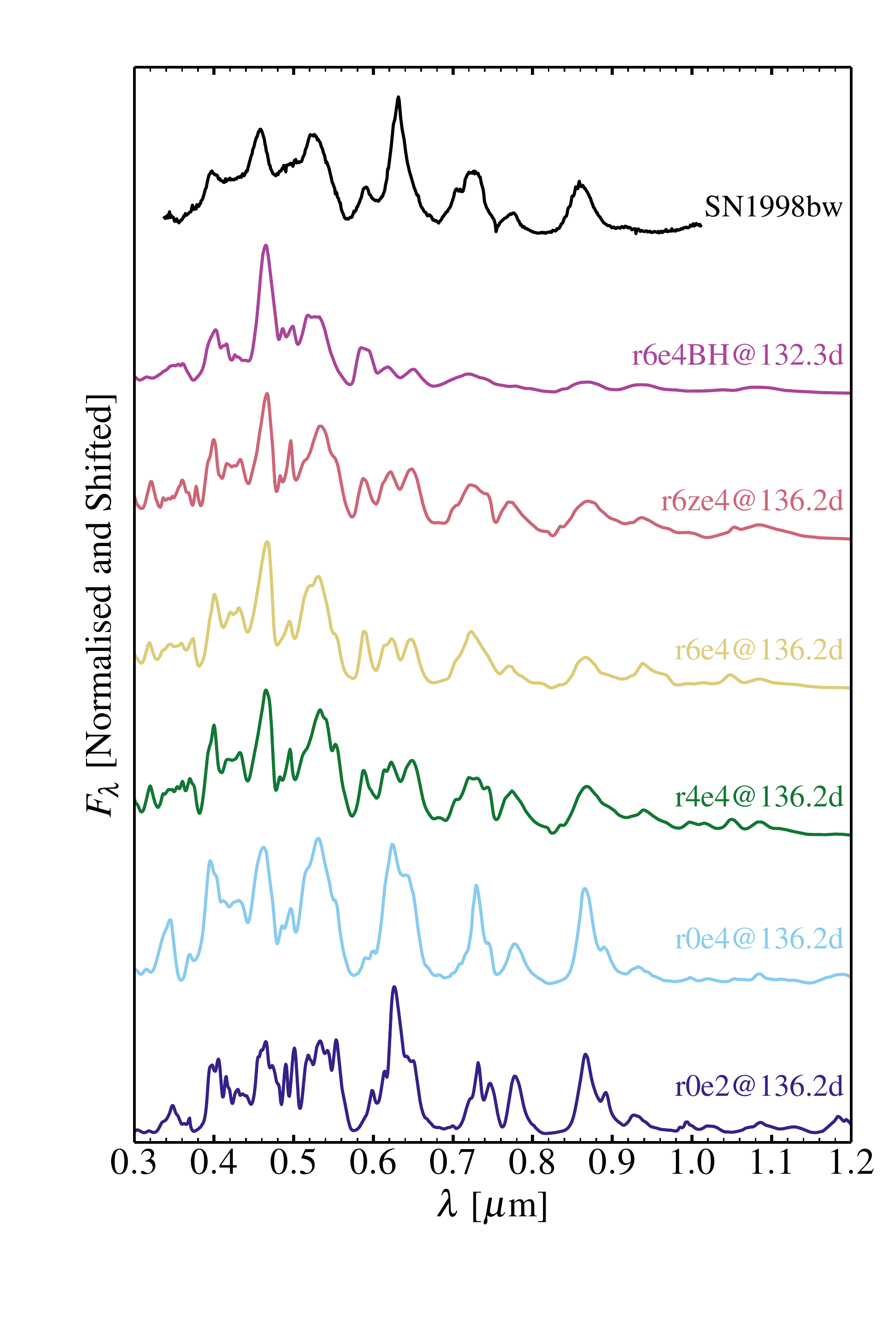,width=8.5cm}
\caption{Optical spectra at early times (left) and late times (right) for
SN\,1998bw (corrected for extinction and redshift; the epochs are 8.4 and 139.4\,d after
the LGRB detection) and for a selection of our models (the post-explosion epoch is indicated
next to each spectrum). Models with a larger $E_{\rm kin}/M_{\rm e}$
come closer to matching the spectrum of SN\,1998bw at early (late) times.
The BH forming SN model r6e4BH matches best the spectral properties of SN\,1998bw at
early times, but models r0e2/r0e4 match better the nebular spectra.
No model matches both the early and late time spectra of SN\,1998bw.
\label{fig_5_135d}
}
\end{figure*}

The full set of models includes r0e2, r0e4,  r4e4, r6e4, r6ze4, and r6e4BH.
Once the explosive nucleosynthesis is over, we mix
\isoni, and only \iso{56}Ni -- we do not mix other species (we do, however, adjust the O mass fraction
to keep the sum of mass fractions equal to
unity).\footnote{In practice, starting at the base of the ejecta, we reset the
\iso{56}Ni mass fraction $X_i$ in each shell $i$ of mass $\delta m_i$ to
$\sum_{k=i}^j X_k \delta m_k/ \Delta m $, where $\Delta m= \sum_{k=i}^j \delta m_k$.
In all models, we use $\Delta m=0.6$\,\msun, except for models r6e4 and r6e4BH for which
we use 0.3\,\msun.
If the sum of mass fractions in mass shell $i$ is greater than unity, we scale all mass fractions so that the sum is unity.
If it is less, we increase the local O mass fraction so that the sum is unity.
In practice, this adjustment is minor.}
Mixing \iso{56}Ni captures the influence of decay heating on the light curve. By not altering
other species, we avoid artificially introducing a microscopic
mixing that probably does not occur in real SN explosions (see, e.g.,
\citealt{jerkstand_87a_11,wongwathanarat_15_3d}).
When mixing is applied to all species, we find little difference
in the SN radiation, probably because these progenitors are quasi-homogeneous
CO cores (this is stark contrast from simulations of SNe II-P; \citealt{lisakov_08bk_17}.

The explosion models computed with \v1d\ are then remapped into \cmfgen\ \citep{HD12} at 1--4\,d
after explosion.
The numerical procedure is the same as used recently in \citet{D15_SNIbc_I,D16_SNIbc_II}.
Because of the large abundance of intermediate-mass elements (IMEs) and Fe-group elements (IGEs)
in these ejecta, we use the same model atoms for metals as used in \citet{D14_tech}, particularly
important to test for the presence of Co\three\ lines \citep{d14_co3} --- this line may be seen and
could be an important diagnostic because SNe Ic-BL are believed to synthesize
a mass of \isoni\ comparable to that of standard SNe Ia \citep{scalzo_ni56_14}.
We show the ejecta composition (for a selection of important species) for models r0e2 and r6e4BH
in Fig.~\ref{fig_init_comp}. Additional ejecta properties are summarised in Table~\ref{tab_sn_prop}.

Our set of simulations was not done to match a specific SN Ic, SN Ic-BL, or LGRB/SN.
Rather, we cover a range of progenitor/ejecta properties and study their impact on observables.
We then confront these results to the photometric and spectroscopic evolution of SN\,1998bw,
using data of \citet{galama_98bw_98}, \citet{mckenzie_schaefer_98bw_99}, and \citet{patat_98bw_01}.
We use an extinction $A_V=$\,0.2\,mag,
a redshift of 0.00867, and a distance modulus of 32.89\,mag \citep{patat_98bw_01}.
We set the explosion time to the time of the LGRB detection (MJD\,50929.4; \citealt{galama_98bw_98}).

\section{Results}
\label{sect_res}

\subsection{Photometry}

Using the $UBVRI$ photometry for both the observations of SN\,1998bw and for our set
of models, we convert these magnitudes to fluxes and integrate to yield a luminosity $L_{UBVRI}$.
Repeating the process from early to late times, we build optical light curves (Fig.~\ref{fig_lum_set_1}).

Our set of six models may be split in two separate groups. The first group contains models r0e2 and r0e4,
which correspond to a massive ejecta with modest amounts of \iso{56}Ni (0.12 and 0.17\,\msun),
and kinetic energies of 4.1 and 12.3$\times$\,\foe. Both models are too faint
to match the peak luminosity of SN\,1998bw.
These models also have a very long rise time to maximum, from
$\sim$\,43\,d (r0e2) down to $\sim$31\,d (r0e4).
Such massive ejecta, typical of what galactic WR stars may produce
(see \citealt{pac_wr_07} for a review), yield SN radiation properties that are incompatible
with the inferred rise times and light curve widths of SNe Ic (see, e.g., \citealt{drout_11_ibc}),
unless they have an explosion energy well in excess of 10$^{52}$\,erg. The exceptional
conditions required to produce such extreme energies
are unlikely to be realised with the frequency at which standard SNe Ic occur. Super-solar
metallicity may be needed to produce the low/moderate-mass WR progenitors required to match
the observed SN Ic properties \citep{georgy_snibc_09,yoon_ibc_15}.

The second set contains models with lower ejecta masses ($\sim$3--8\,\msun),
large explosion energies ($\sim$\,12$\times$\,10$^{51}$\,erg),
and large \iso{56}Ni masses (0.3--0.7\,\msun).
The higher the \iso{56}Ni mass, the greater the bolometric luminosity at maximum
(compare model r6ze4 to model r6e4).
The larger the $M_{\rm e}$, the longer the rise time to maximum and the broader the light curve
(compare model r6ze4 to r6e4BH).
The greater the $E_{\rm kin}/M_{\rm e}$, the faster the decline rate at late times.
These trends are the same as those obtained in a previous grid of models of SNe IIb/Ib/Ic
\citep{D15_SNIbc_I,D16_SNIbc_II}.

The model that best matches the light curve of  SN\,1998bw at early times
is the BH forming model. The model peaks at 20\,d, only 5\,d later than observed,
but it is also a little too faint.
The model light curve width is comparable to that of SN\,1998bw. However, the low ejecta mass
(which allows the very fast rise to maximum)  causes an enhanced $\gamma$-ray
escape at late times, producing a low SN luminosity.
Models r4e4 and r6ze4 are less discrepant at late times but they overestimate the rise
time to maximum and the light curve width.
Model r6e4 is too faint at most times --- it contains too little \iso{56}Ni initially.

In contrast, model r0e2 fits poorly the epochs around maximum but it follows closely
the optical luminosity of SN\,1998bw at nebular times, despite its modest ejecta energetics and \iso{56}Ni mass
of only 0.122\,\msun.  The lower decay power yields a much higher luminosity than in model r6e4BH
because of the near complete trapping of $\gamma$-rays.

Hence, we corroborate the results of \citet{woosley_98bw_99}. To come close
to matching the early-time light curve, a spherically-symmetric model needs
a huge ejecta kinetic energy well in excess of 10$^{52}$\,erg, 0.5\,\msun\ of \isoni,
and a low/moderate ejecta mass.
But, as emphasized by \citet{maeda_98bw_03}, these highly energetic ejecta tend
to yield a luminosity at late times that is too faint or  decreases over time with the wrong rate.

\subsection{Spectroscopy}

We now discuss the spectral evolution of our set of models from pre-maximum to the
nebular phase, and compare to the observations of SN\,1998bw.
Contrary to all previous studies that focused on snapshots (see, e.g., \citealt{nakamura_98bw_01}),
this work  presents the first spectral model sequences based on non-LTE time-dependent
radiative transfer simulations for SNe Ic-BL.

\subsubsection{Early times}

Figure~\ref{fig_5_135d} compares the optical spectra for our model set at $\sim$8\,d
and $\sim$135\,d after explosion with the contemporaneous observations of SN\,1998bw.
At early times, all models fail to reproduce the width of the observed absorption features,
despite their large $E_{\rm kin}/M_{\rm e}$ ratio, even for model r6e4BH.
The model spectra have both narrower line features and stronger absorption troughs than SN\,1998bw.
Hence, this suggests an even  larger energy per unit mass is needed to match the observations.

Our models are also too red, except for the low-metallicity model r6ze4, which comes close
to matching the optical colours of SN\,1998bw at 8\,d. There is a complicated combination of effects
since model r6ze4 has a low primordial metallicity but this model is also very metal
rich owing to the secular and explosive nucleosynthesis that took place. At 8\,d, Ni/Co
do not greatly pollute the outer ejecta and they have not yet decayed to raise sizeably the Fe abundance.
Hence, the Fe mass fraction is down by a factor of ten compared to solar in the outer ejecta.
At early times, a lower primordial metallicity can therefore produce a bluer SN.

As illustrated for models r0e2 and r6e4BH in the appendix
(Figs.~\ref{fig_r0e2_ladder_1}--\ref{fig_r6e4bh_ladder_2}),
the early-time optical spectra show neutral or once ionized lines of IMEs.
The larger the ratio $M$(\iso{56}Ni)/$M_{\rm e}$, the larger the ionisation in
the spectrum formation region.
In model r0e2, the spectra show the presence of lines from C\one\ (red part of the optical),
O\one, Na\one\,D, Mg\one\, Ca\two, together with strong blanketing from Fe\one\
shortward of 5500\,\AA. There is no sizeable contribution from Si\two\,6355\,\AA.
In model r6e4BH, the ionisation is much higher and the expansion rate in the spectrum
formation is larger. We therefore see broad lines of C\one, O\one, Si\two, Ca\two, and Fe\two.

While all the features seen in model r6e4BH are compatible with those observed in the spectrum
of SN\,1998bw at 8.4\,d, they are too narrow.
If we were to scale this model to the 10\,\msun\ ejecta required by \citet{iwamoto_98bw_98}, this
would correspond to a 4.6$\times 10^{52}$\,erg, and the lines would still be too narrow.
Furthermore, this energy approaches the theoretical upper limit of $\sim$\,10$^{53}$\,erg, which
is the gravitational binding energy of a typical neutron star.

\begin{figure}
\epsfig{file=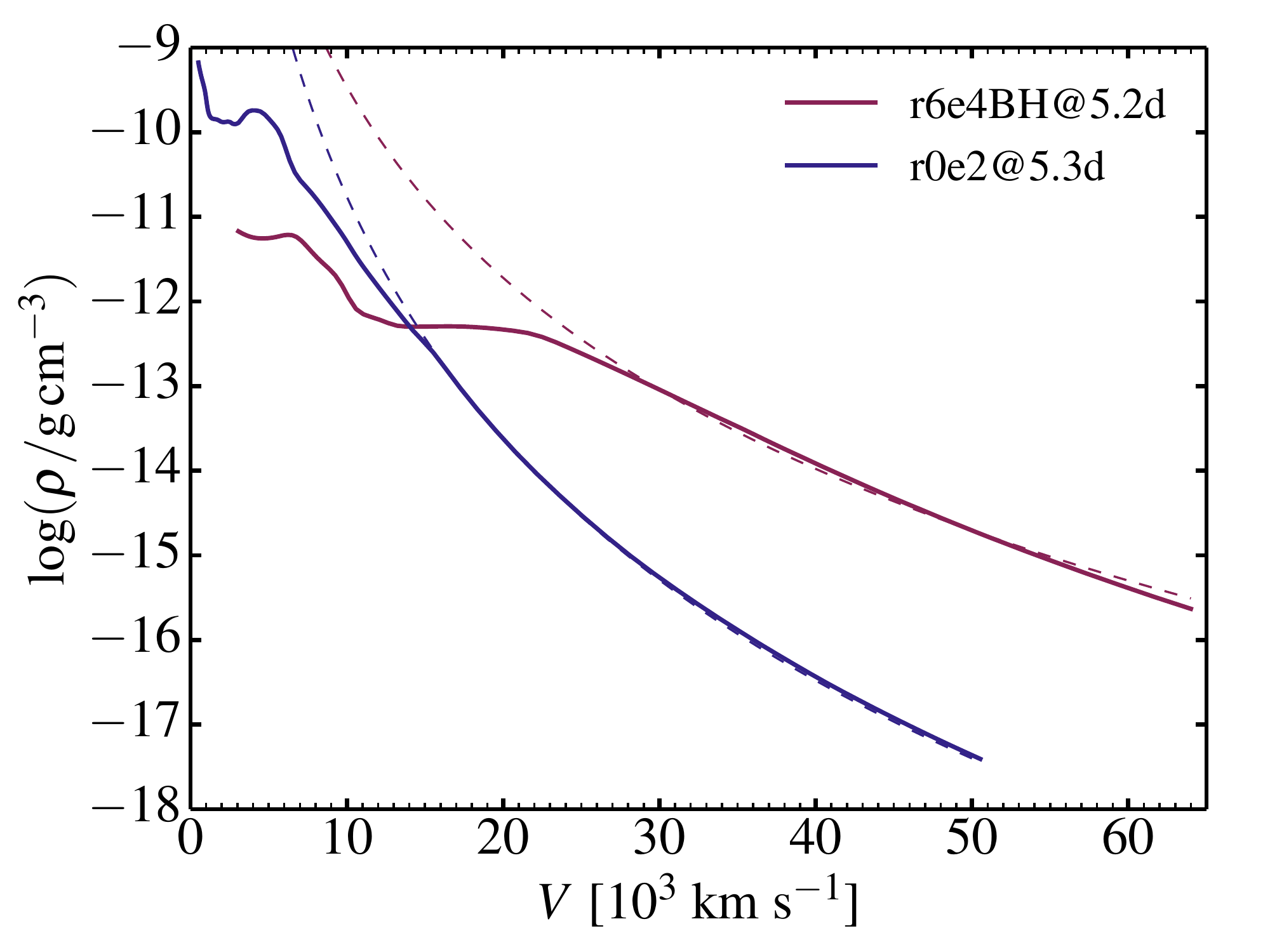,width=8.5cm}
\caption{Comparison of the density versus velocity for models r6e4BH and r0e2 at $\sim$\,5\,d after explosion.
Dashed lines represent fits to the density distribution using the expression
$\log \rho(V) = \log \rho(V=V_0) + N_\rho \log (V_0/V)$, where $V_0=$\,30,000\,\kms\ and $N_\rho=$\,7.5
for model r6e4BH and $V_0=$\,20,000\,\kms\ and $N_\rho=$\,9.5 for model r0e2.
The greater the energy of the explosion, the more material is present at large velocity.
In spherical symmetry, radiation hydrodynamics cannot produce high density
regions at both low and high velocity.
However, by adopting an axisymmetric configuration it is possible, for example to have dense material at
high velocities in the  polar direction, and at low velocity in the equatorial regions.
\label{fig_rho_vel}
}
\end{figure}

\cite{nakamura_98bw_01} modelled the early time spectra of SN\,1998bw and found similar
properties for the spectrum formation at early times, including very fast expansion (with absorption
extending out to 60,000\,\kms\ in some lines, a composition dominated by intermediate mass elements).
Their line identifications are identical to ours around the epoch of maximum.
They note that the line absorptions are better fitted when invoking a flatter density distribution at large
velocity (they argue for a power law with a density exponent of $-6$, while we have a value of $-7.5$
in model r6e4BH; Fig.~\ref{fig_rho_vel}).

None of our models shows the strong optical He\one\ lines characteristic of SNe Ib -- all our
models produce SNe Ic (or SNe Ic-BL), although they all contain some He.
The total He mass is in the range 0.18--1.4\,\msun,
with a surface He mass fraction of 20\% (model r0, r4, r6) and 98\% (model r6z).
However, model r6e4BH shows the presence of He\one\,10830\,\AA, although the line
is quite weak at this early time.
Hence, He deficiency is not mandatory for producing a SN Ic or a SN Ic-BL (the same result is
obtained for the SN Ic ejecta models 5p11Ax1/5p11Ax2 in \citealt{D15_SNIbc_I}).

\subsubsection{Late times}

In contrast to early times, our model set produces a larger diversity in spectral
appearance at late times (Fig.~\ref{fig_5_135d}, right panel).
This results from the differences in ionisation (greater in ejecta with
a larger $M$(\iso{56}Ni)/$M_{\rm e}$ ratio) and composition.

Models with a low $M$(\iso{56}Ni)/$M_{\rm e}$ ratio (r0e2 and r0e4) show
a spectrum dominated by O\one\ lines at 6300--6363\,\AA\ (forbidden doublet)
and 7774\,\AA, Na\one\,D, the Ca\two\,7300\,\AA\ doublet and the
Ca\two\ near-IR triplet around 8500\,\AA, and lines of Fe\two.
In the blue part of the optical, we see the overlapping contributions of forests of lines from Ti\two, Fe\two,
as well as strong absorptions, for example from Ca\two\,H\&K.

In contrast, model r6e4BH shows a dominance of IGEs, with the strong Fe\two, Fe\three, and Co\three\ lines
(Fig.~\ref{fig_5_135d}), and a spectral energy distribution that peaks in the blue part of the optical.
There is no emission associated with O\,\one\,6300--6364\,\AA, but we see a sizeable
emission in [Co\three]\,5888\,\AA, as also observed in SNe Ia \citep{d14_co3}.
Overall, this model spectrum resembles a Type Ia spectrum at late times, which is not surprising
given the model properties ($M_{\rm e}=$\,2.55\,\msun, $M$(\iso{56}Ni)$=$\,0.435\,\msun).

Model r6e4BH, despite its suitability for the early time properties of SN\,1998bw, is
strongly discrepant at late times. In contrast, model r0e2 matches quite closely
the spectral properties of SN\,1998bw, even though it corresponds to the explosion
of a massive progenitor yielding a  low/moderate energy and a low \iso{56}Ni mass
($M_{\rm e}=$\,9.69\,\msun, $E_{\rm kin}=$\,4.12$\times$\,10$^{51}$\,erg,
$M$(\iso{56}Ni)$=$\,0.122\,\msun).

From the above discussion we see that while our models reproduce some of the features
of the optical light curve and spectra of SN\,1998bw at specific epochs,
all models also show significant discrepancies when compared to observations at all epochs.

\section{Discussion and conclusions}
\label{sect_conc}

We have presented physically-consistent explosion models
for rotating/non-rotating C/O-rich WR stars. Starting at 1--4\,d after explosion,
we follow each model with the 1-D time-dependent radiative-transfer code \cmfgen,
allowing for thermal-heating and non-thermal processes associated with the
radioactive decay of \iso{56}Ni and \iso{56}Co. Our approach allows, for the first
time, the simultaneous computation of bolometric, multi-band light curves and spectra
for models of SNe Ic-BL.
While our modelling confirms some results from previous work,
it also conflicts with numerous recent studies of LGRB/SNe.

We find that a high energy
explosion yielding a high $E_{\rm kin}/M_{\rm e}$ can approximately reproduce
the light curve properties around maximum (rise time, peak brightness,
light curve width) as well as the very broad optical lines (associated with Fe\two, Si\two, or Ca\two)
observed in SN\,1998bw.
But this model fails at late times because it underestimates the luminosity, and
yields spectra dominated by metal lines without the presence of, e.g., O\one\,6300--6364\,\AA.
Conversely, we find that a lower energy explosion yielding a lower $E_{\rm kin}/M_{\rm e}$,
which fails at early times, produces a satisfactory match to both the observed luminosity and
optical spectra at nebular times.

\citet{maeda_98bw_03} suggested that the observed light curve of SN\,1998bw can be reconciled
by invoking a highly energetic ejecta with a distribution of \isoni\ in two concentric
shells at low and high velocity, and a high-density inner ejecta.
In our work, and as illustrated in Fig.~\ref{fig_rho_vel},
we cannot reproduce this configuration because a high energy explosion model (e.g., r6e4BH)
has little or no material at low velocity, while a  low energy explosion, has little or no material
at high velocity (e.g., r0e2) --- the ejecta structure with two-concentric and spherical shells
proposed by \citet{maeda_98bw_03} is not hydrodynamically feasible.
In other words, a spherically-symmetric explosion model cannot have simultaneously a large
density at low and high velocity. A simple way out of this conundrum is to argue that
the explosion at the origin of SN\,1998bw is asymmetric, yielding, for example,  a prolate (oblate)
density distribution along the pole (equator). Such a morphology has been proposed by
\citet{maeda_neb_06,maeda_neb_08} based on modelling of nebular phase spectra.
We demonstrate here that
evidence for asymmetry is in fact visible at all times, both in the photometry and in the spectra.

Instead of a model with two concentric shells of \iso{56}Ni located at low and high velocities,
the ejecta may be axisymmetric with fast
material (high energy-per-unit-mass with high relative abundance of \isoni\ with respect to
other species) limited to a small/moderate solid angle around our line of sight
(and possibly 180\,deg away
from our line of sight), and slower material (lower energy-per-unit-mass with a lower relative
abundance of \isoni\ with respect to other species) lying along lower latitudes.
This hybrid scenario is qualitatively similar to the axisymmetric explosions of  \citet{MacFadyen_collapsar_99} or
\citet{maeda_2d_98bw_02}.

\begin{figure*}
\epsfig{file=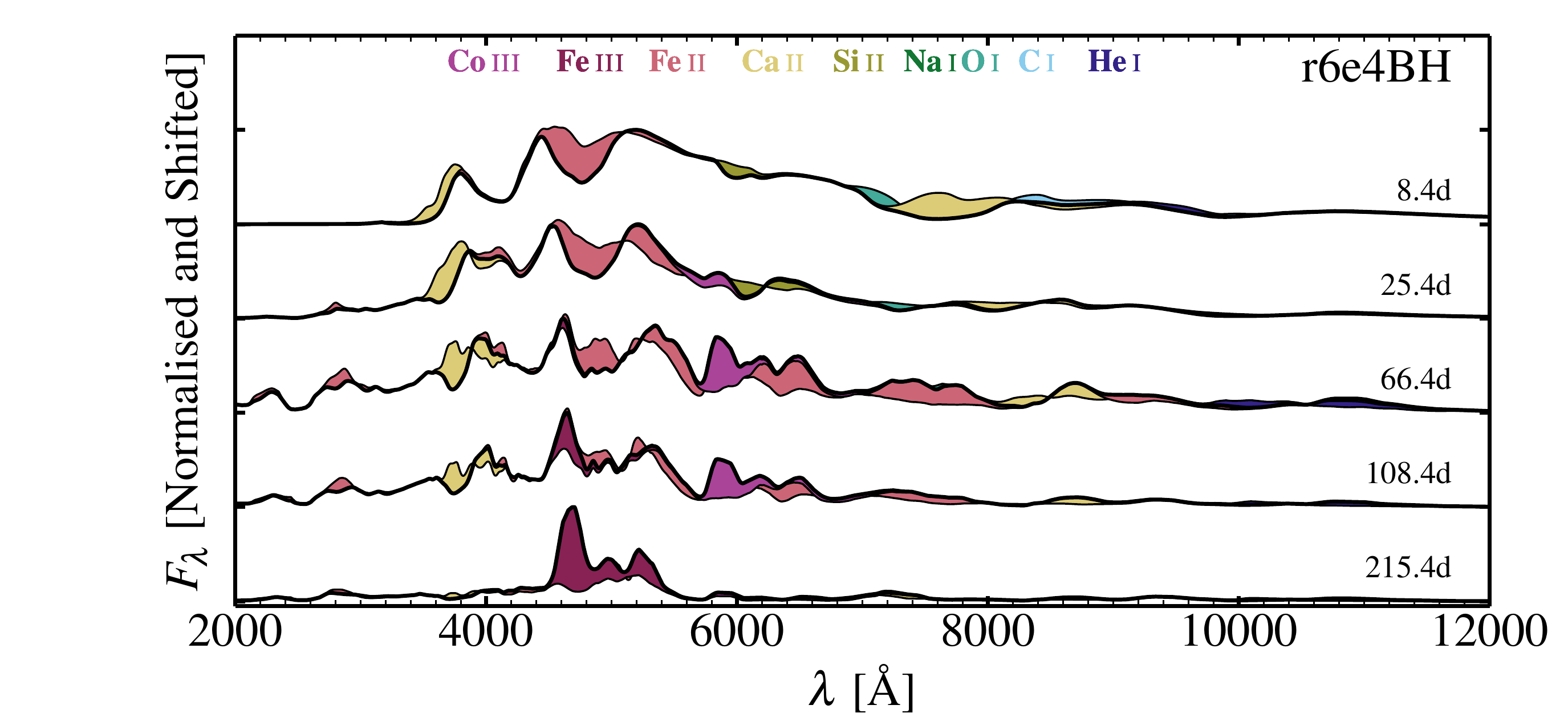,width=16.2cm}
\epsfig{file=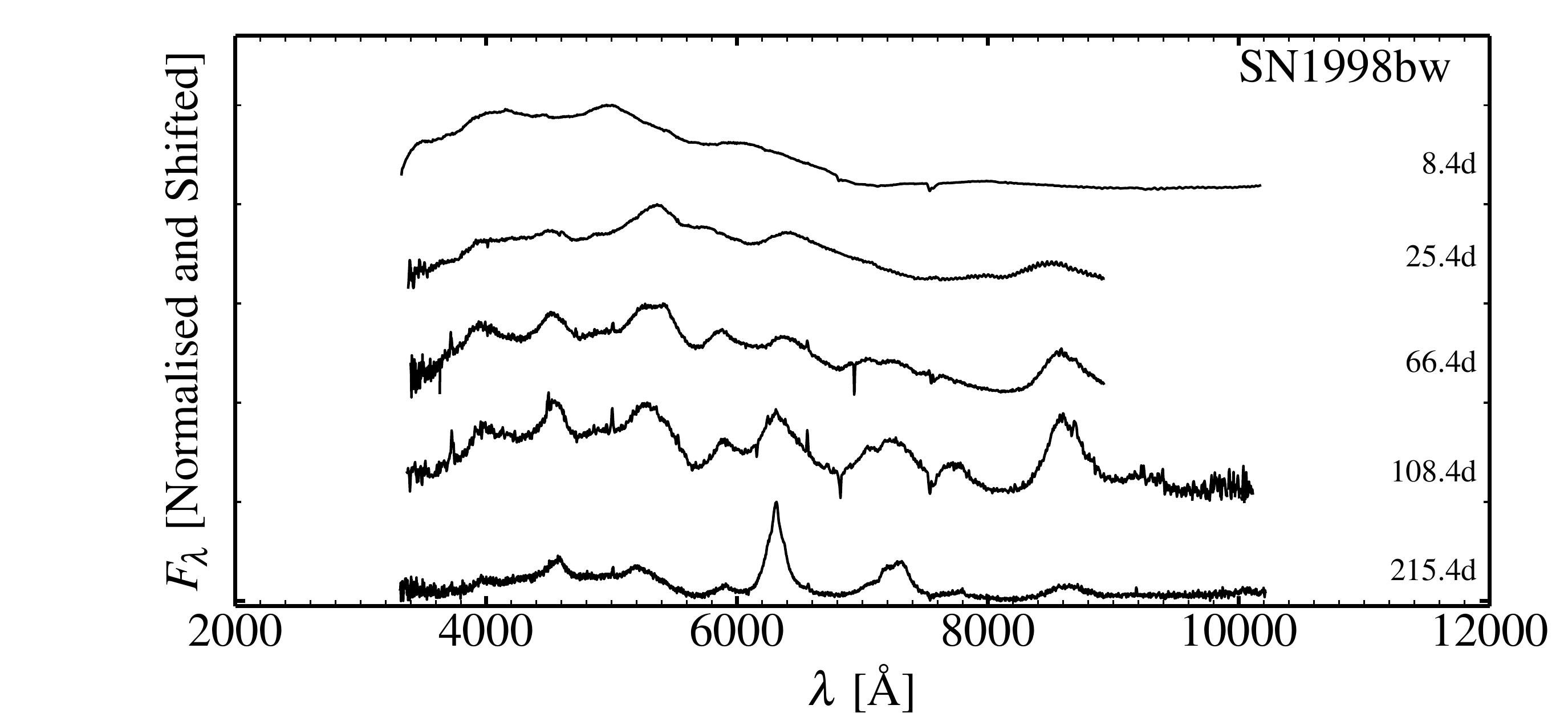,width=16.2cm}
\epsfig{file=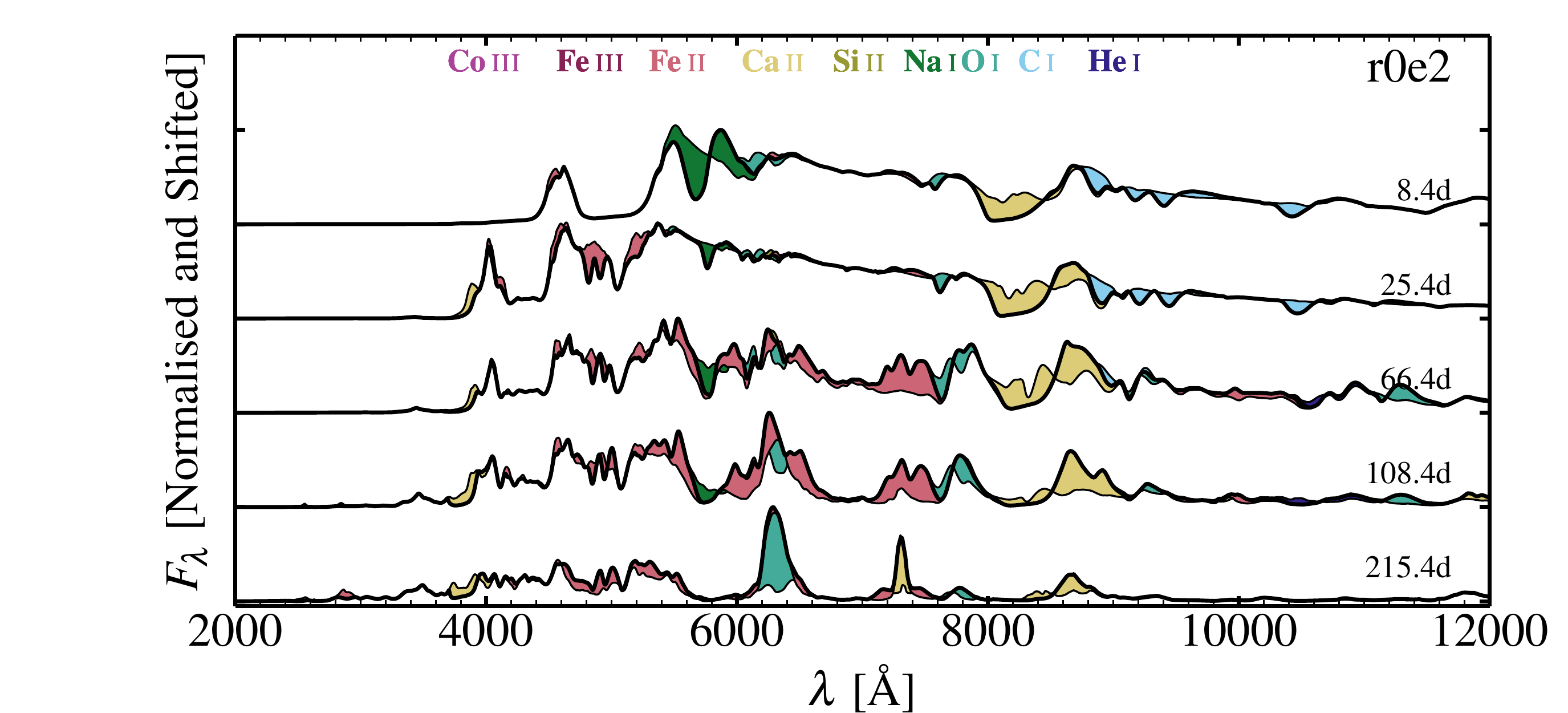,width=16.2cm}
\caption{Multi-epoch spectra of SN\,1998bw (center; corrected for redshift and extinction),
and of models r6e4BH (top) and r0e2 (bottom). Times are given since MJD\,50929.4 for SN\,1998bw
and since explosion for the models. We shade the model flux associated with bound-bound
transitions of specific ions (although instructive, this procedure only works accurately
in the absence of line overlap).
Neither model matches the whole evolution of SN\,1998bw.
However, the higher  energy model r6e4BH more closely resembles SN\,1998bw
at early times while the lower energy model (r0e2) matches better at late times.
\label{fig_montage_asym}
}
\end{figure*}

For the fast material along the pole, the ejecta properties could correspond to model r6e4BH,
but limited to a small solid angle. At early times, the ejecta is optically thick and we are mostly
sensitive to material along our line of sight.
Consequently, the total mass of the very fast material and of its \isoni\ content are reduced
(perhaps to a tenth of the masses in the spherical ejecta model r6e4BH, if this region occupies a tenth
of the volume of this spherical model). The radiation from this material is not beamed by a relativistic
effect, but it is non-isotropic and biased in the radial direction along which the photon mean free path
is  the greatest. Thus at early times, the spectra would form primarily in the optically-thick column
of fast-moving material along our line of sight, resembling the properties of our (1-D) model r6e4BH.
At late times, the \isoni\ at large velocity contributes negligibly to the light curve
(because of $\gamma$-ray escape) and also because the corresponding mass is lower than
the 0.5\,\msun\ value of the corresponding spherical model (how much lower will depend on
the opening angle for the fast material and may vary from case to case).

This model reproduces roughly the salient features seen in the early-time spectra of SN\,1998bw,
with very broad Fe\two\ and Si\two\ lines, a broad feature in the red from the blended absorption
of O\one-Ca\two.
The key here is that the huge energy and huge \isoni\ inferred to match SN\,1998bw
arise when spherical symmetry is assumed. The early-time light curve and spectra
may instead be reproduced by an asymmetric ejecta, with a much lower energy and \isoni\
mass for the fast moving material because it is limited to a small solid angle.

The bulk of the mass located at lower velocities could correspond to an ejecta mass of 10\,\msun\
with an energy of a few \foe\ and 0.1\,\msun\ of \isoni. Our model r0e2 might be a rough representation
of this inner ejecta in SN\,1998bw.
Only 0.1\,\msun\ of \iso{56}Ni may be needed to reproduce the luminosity of SN\,1998bw at late times
(Fig.~\ref{fig_lum_set_1}) because the denser, slower, and more massive ejecta leads to the more efficient
trapping of $\gamma$-rays.
The large O mass and the low ionisation conditions lead to the production of a strong
O\one\,6300--6364\,\AA\ line, as observed. In contrast, model r6e4BH shows only lines of IGEs at late times,
and no O\one\ line emission. With both models combined, we can explain the simultaneous presence of
Fe\three, Co\three, and O\one\ lines at nebular times.
Figure~\ref{fig_montage_asym} illustrates how an asymmetric model may reproduce
the spectral observations of SN\,1998bw, resembling model r6e4BH at $\lesssim$\,70\,d
and model r0e2 later on.

Allowing for asymmetry, it is possible for LGRB/SN progenitors to be massive WR stars.
Unlike in the 1-D $\sim$\,5\,\msun\ ejecta model of \citet{woosley_98bw_99},
only a small fraction of the ejecta mass is rich in \isoni, contributing around bolometric maximum,
while the longer-time light curve is powered by a modest mass of \isoni\ but tied to a more massive
ejecta.
In this context, the mass-loss/angular-momentum-loss problem
that plagues LGRB/SN progenitors no longer holds. Furthermore, if we argue for the formation
of a massive BH, massive WR stars (as massive as models 35OB-35OC in \citealt{WH06})
are suitable for producing a LGRB/SN. This model still requires fast-rotating progenitors evolved
at low metallicity \citep{hirschi_rot_04,yoon_grb_05,georgy_snibc_09},
but there is no longer the need to reach down to final WR star masses as low as 5\,\msun.
The possibility of a fast-rotating massive WR progenitor for LGRB/SNe
means that the proto-magnetar model is not the only viable model.
\cite{mazzali_pm_14} promote the proto-magnetar model of
LGRB/SNe because their energetics is comparable to the initial energy of a millisecond-period
magnetar. This may be fortuitous. For example, standard core-collapse and thermonuclear SNe
have a similar explosion  energy but their explosion mechanism is entirely different and their
progenitors come from distinct stellar populations.
\citet{wang_pm_17} claim that the early peak and late-time slow decline in the light curves
support the proto-magnetar model because no other model can match this feature.
However, our model shows this is clearly not the case.
The early peak and late-time slow decline in the light curves
are indicative of ejecta asymmetry, and our models of spectra at the photospheric
and nebular phases confirm this.

Spherically symmetric models of LGRB/SN light curves and spectra are characterized
by energies up to half the binding energy of a neutron star. By invoking asymmetry, LGRB/SNe
may have ejecta energies
$\lesssim$\,10$^{52}$\,erg, thus more compatible with expectations of what can be produced in
fast-rotating progenitors. At cosmological redshifts, we tend to detect events in which the bulk of the
energy is injected along our direction (as high-energy radiation), and then infer a large
``isotropic'' luminosity, while we fail to detect those in which the energy is deposited away
from our line of sight, even though they may have the same energetics.
LGRB/SNe are extremely rare events relative to the ensemble of core-collapse SNe
(see, e.g., \citealt{podsiadlowski_grb_04}), so the reduced \iso{56}Ni mass brought
in by considering asphericity would have little impact on the chemical enrichment of the Universe
(the higher mass ejecta would however favor a greater release of O from lower latitudes).

Even today, the majority of analyses on LGRB/SNe and standard SNe Ic are based
on simplistic arguments for the LC \citep{drout_11_ibc, dhelia_13dx_15, prentice_ibc_16}.
\citet{volnova_13dx_17} performed detailed radiation-hydrodynamics simulations of LGRB/SN\,2013dx,
but their modeling is 1-D and they focus on the early-time radiation only.
When applied to the asymmetric ejecta of LGRB/SNe, such studies can lead to
systematic errors in the determination of ejecta parameters.

The large mass budget inferred for SN\,1998bw makes it possible to argue for BH formation.
As discussed in \citet{D12_BH}, the final density structure of the 12-16\,\msun\ models of \citet{WH06}
proposed as LGRB/SN progenitors is similar to that of standard non-rotating 15\,\msun\ RSG stars.
The latter are expected to produce garden-variety SNe II-P, so it is unlikely that the former could
produce the extreme properties necessary to make LGRB/SNe, which are also very rare events.
On the other hand, the massive WR models 35OB/35OC of \cite{WH06}, with their massive Fe cores
and high compactness, now appear suitable to form a collapsar and an LGRB/SN like SN\,1998bw.

\begin{acknowledgements}

LD acknowledge financial support from the European Community through an
International Re-integration Grant, under grant number PIRG04-GA-2008-239184,
and from ``Agence Nationale de la Recherche" grant ANR-2011-Blanc-SIMI-5-6-007-01.
DJH acknowledges support from  NASA theory grant NNX10AC80G,
and SCY acknowledges support from the Korea Astronomy and Space Science Institute
under the R\&D program (Project No. 3348-20160002) supervised by the Ministry of Science,
ICT and Future Planning.
This work utilized computing resources of the mesocentre SIGAMM,
hosted by the Observatoire de la C\^{o}te d'Azur, Nice, France.

\end{acknowledgements}

\appendix

\section{Line identifications in model \lowercase{r0e2} at multiple epochs}

In this section, we present a montage of spectra for model r0e2 that illustrates
the bound-bound contributions  from selected ions at four consecutive epochs.

\begin{figure*}
\epsfig{file=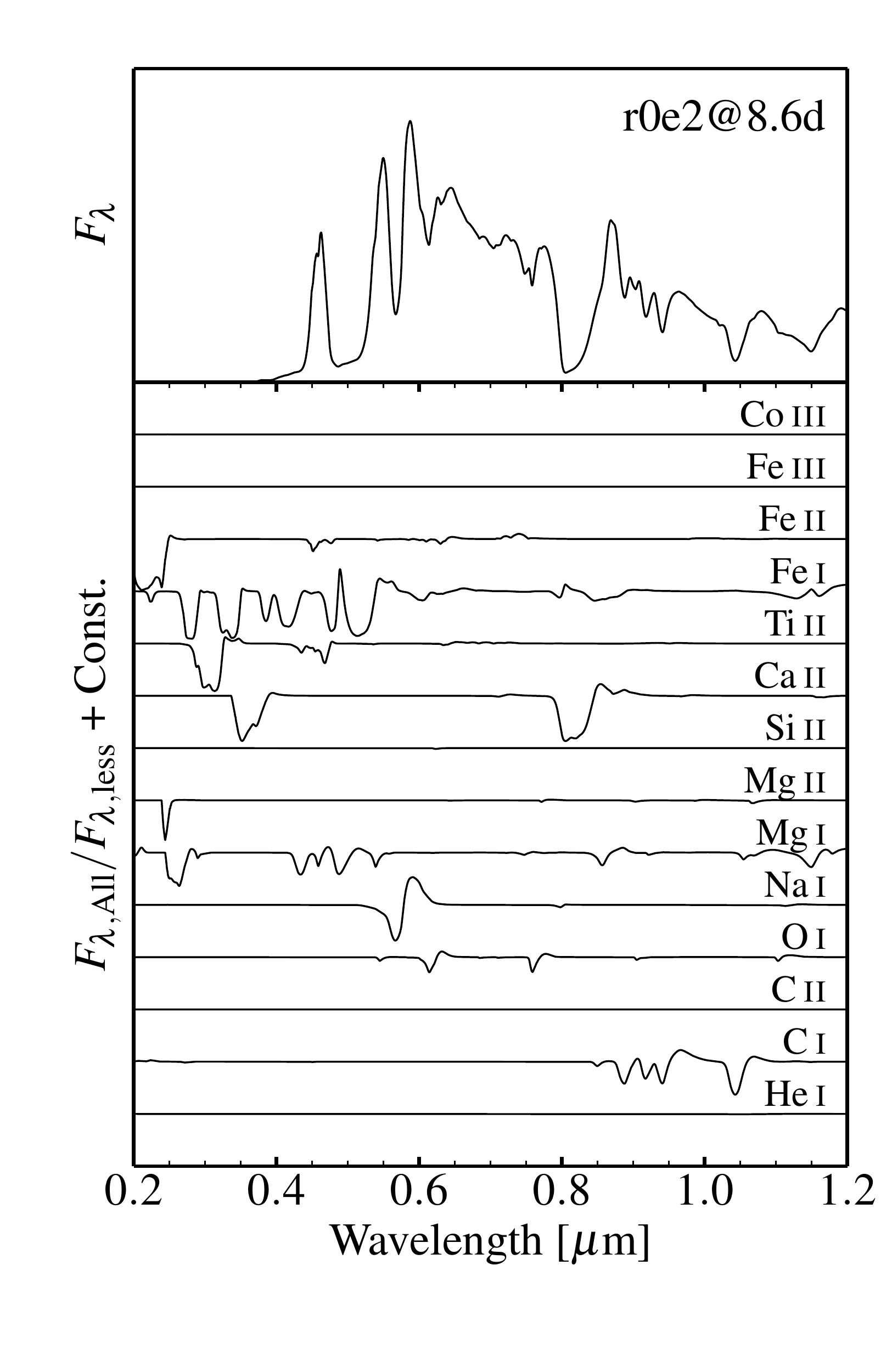,width=7.1cm}
\epsfig{file=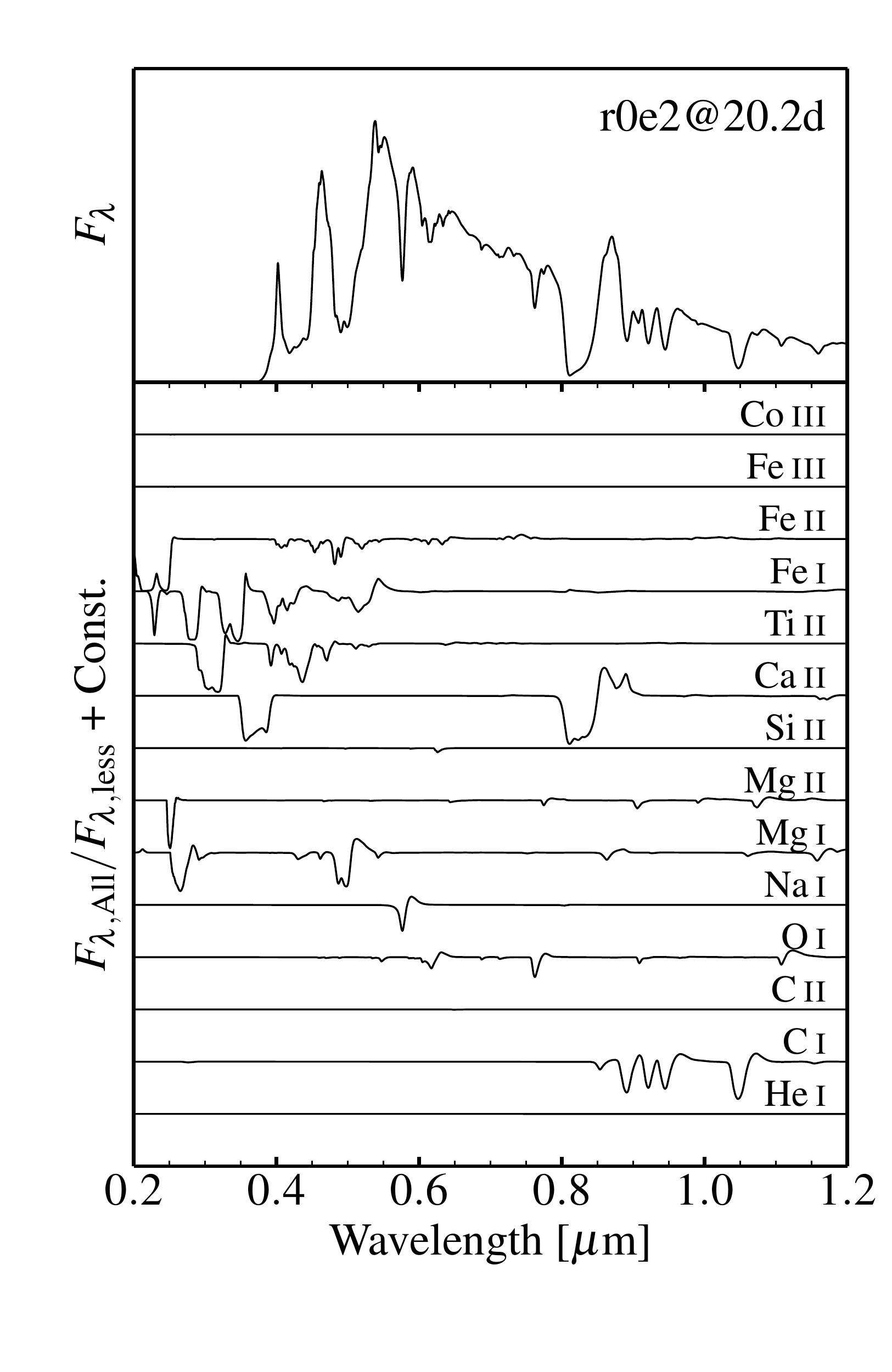,width=7.1cm}
\caption{Montage of spectra for model r0e2 for post-explosion times of 8.6 (left) and 20.2\,d (right).
In the top part of each panel, we show the total spectrum,
while below, we stack the ratio between the full synthetic spectrum
($F_{\lambda,{\rm All}}$) and the synthetic spectrum computed by ignoring the bound-bound transitions
of a given ion ($F_{\lambda,{\rm less}}$; as indicated by the label on the right).
\label{fig_r0e2_ladder_1}
}
\end{figure*}

\begin{figure*}
\epsfig{file=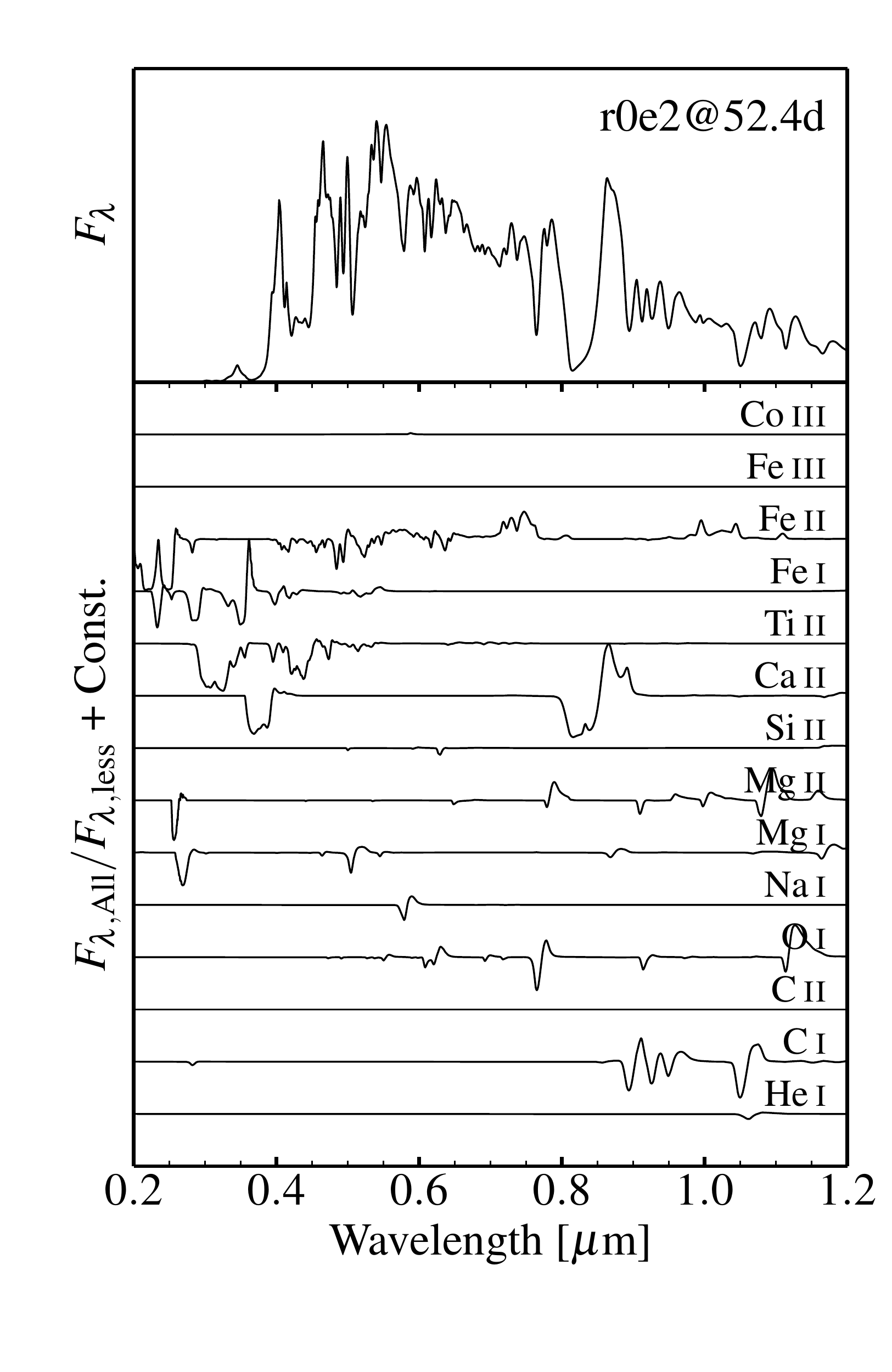,width=7.1cm}
\epsfig{file=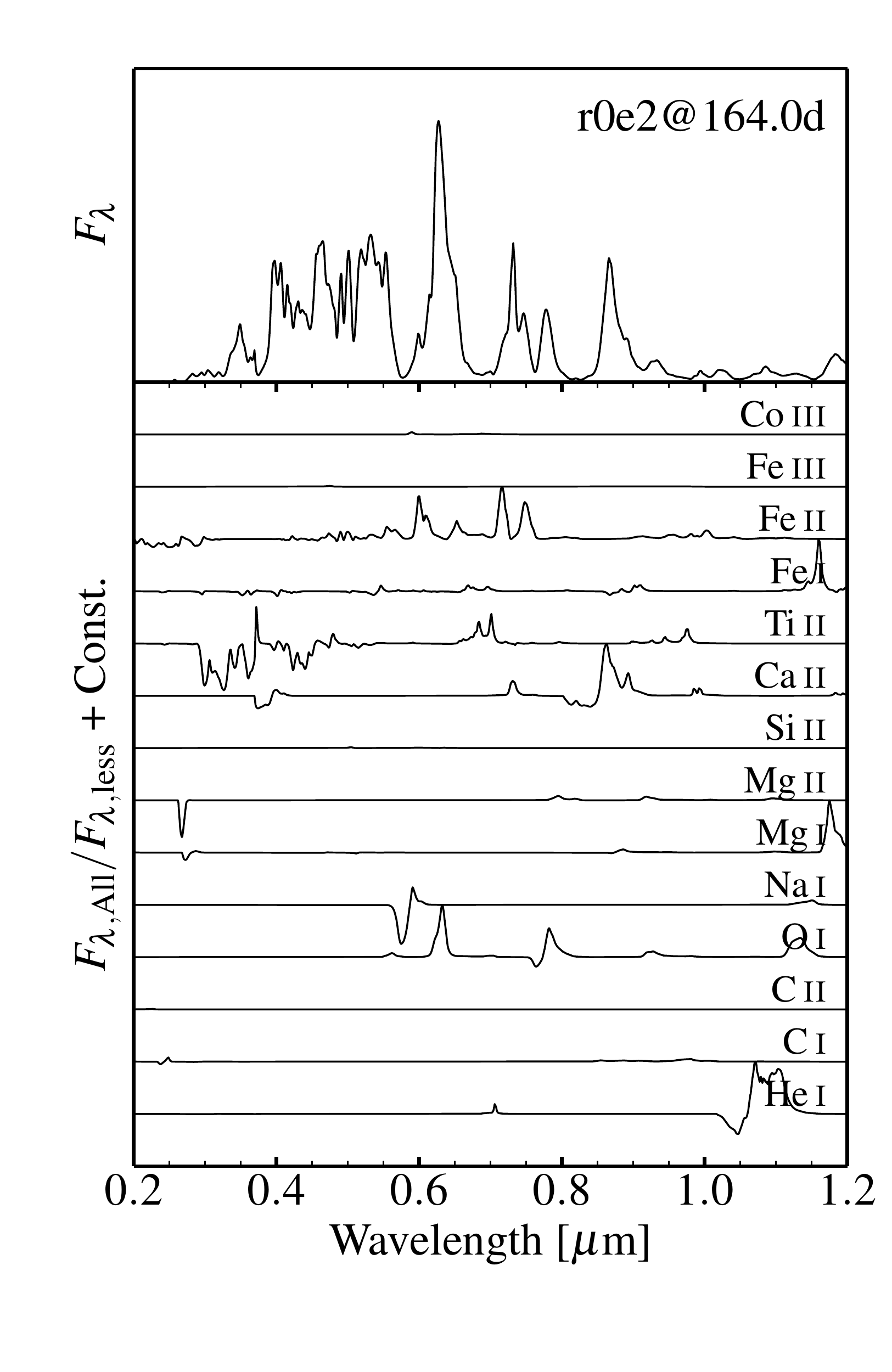,width=7.1cm}
\caption{Same as Fig.~\ref{fig_r0e2_ladder_1}, but now for post-explosion times
52.4 (left) and 164.0\,d (right).
\label{fig_r0e2_ladder_2}
}
\end{figure*}

\section{Line identifications in model \lowercase{r6e4}BH at multiple epochs}

In this section, we present a montage of spectra for model r6e4BH that illustrates
the bound-bound contributions  from selected ions at four consecutive epochs.

\begin{figure*}
\epsfig{file=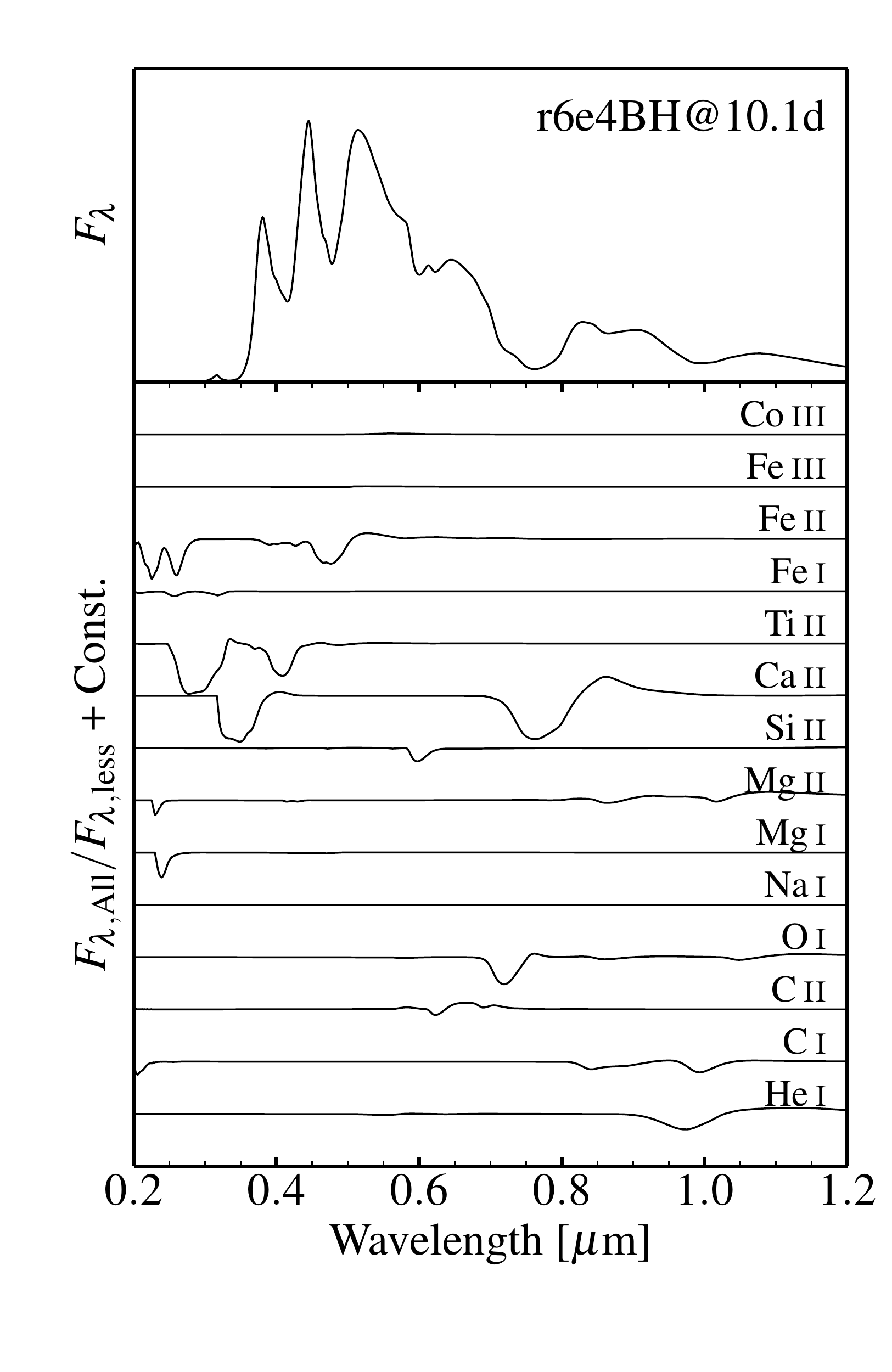,width=7.1cm}
\epsfig{file=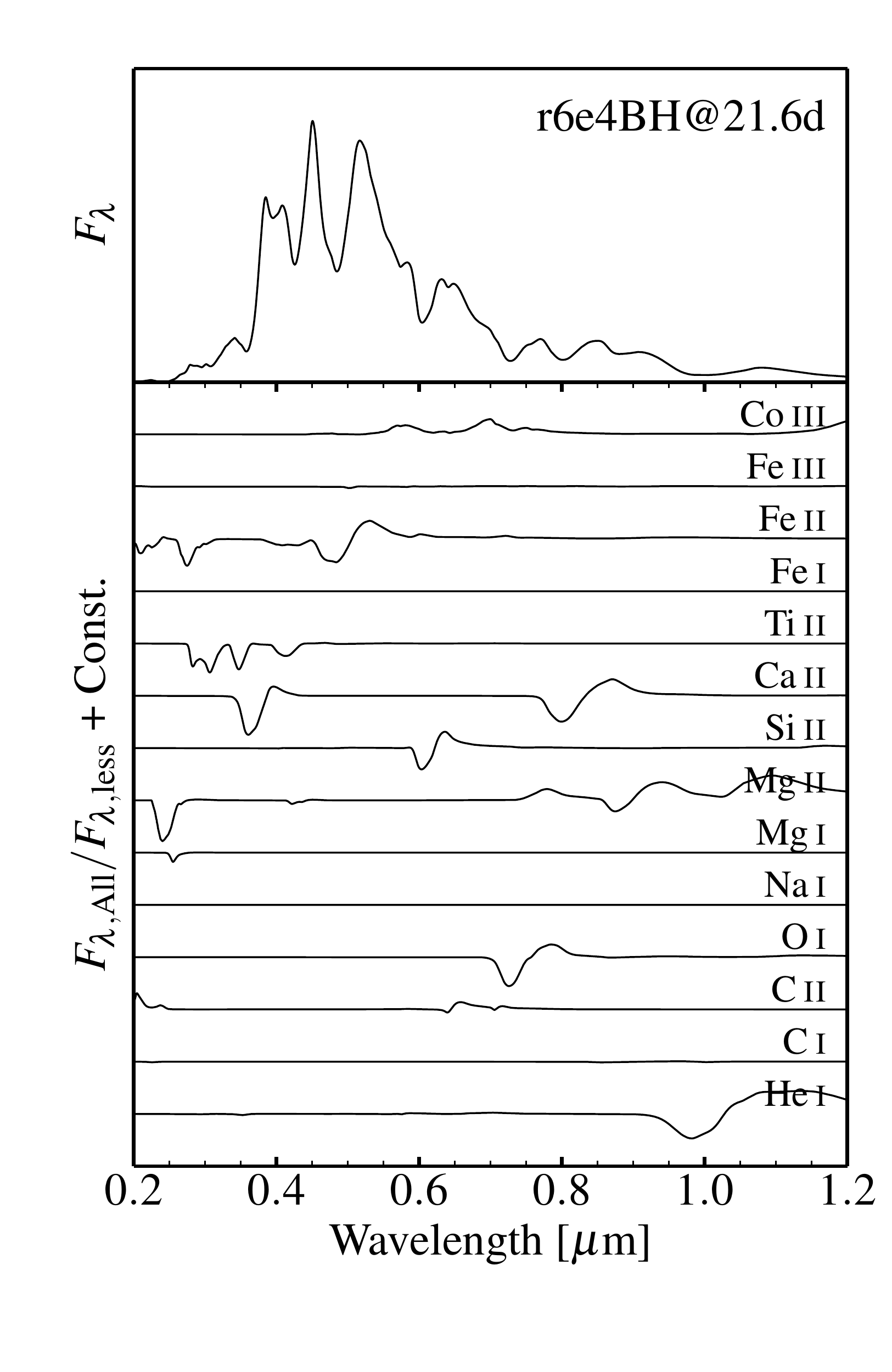,width=7.1cm}
\caption{Same as Fig.~\ref{fig_r0e2_ladder_1}, but now for model r6e4BH at post-explosion times
of 10.1 (left) and 21.6\,d (right).
\label{fig_r6e4bh_ladder_1}
}
\end{figure*}

\begin{figure*}
\epsfig{file=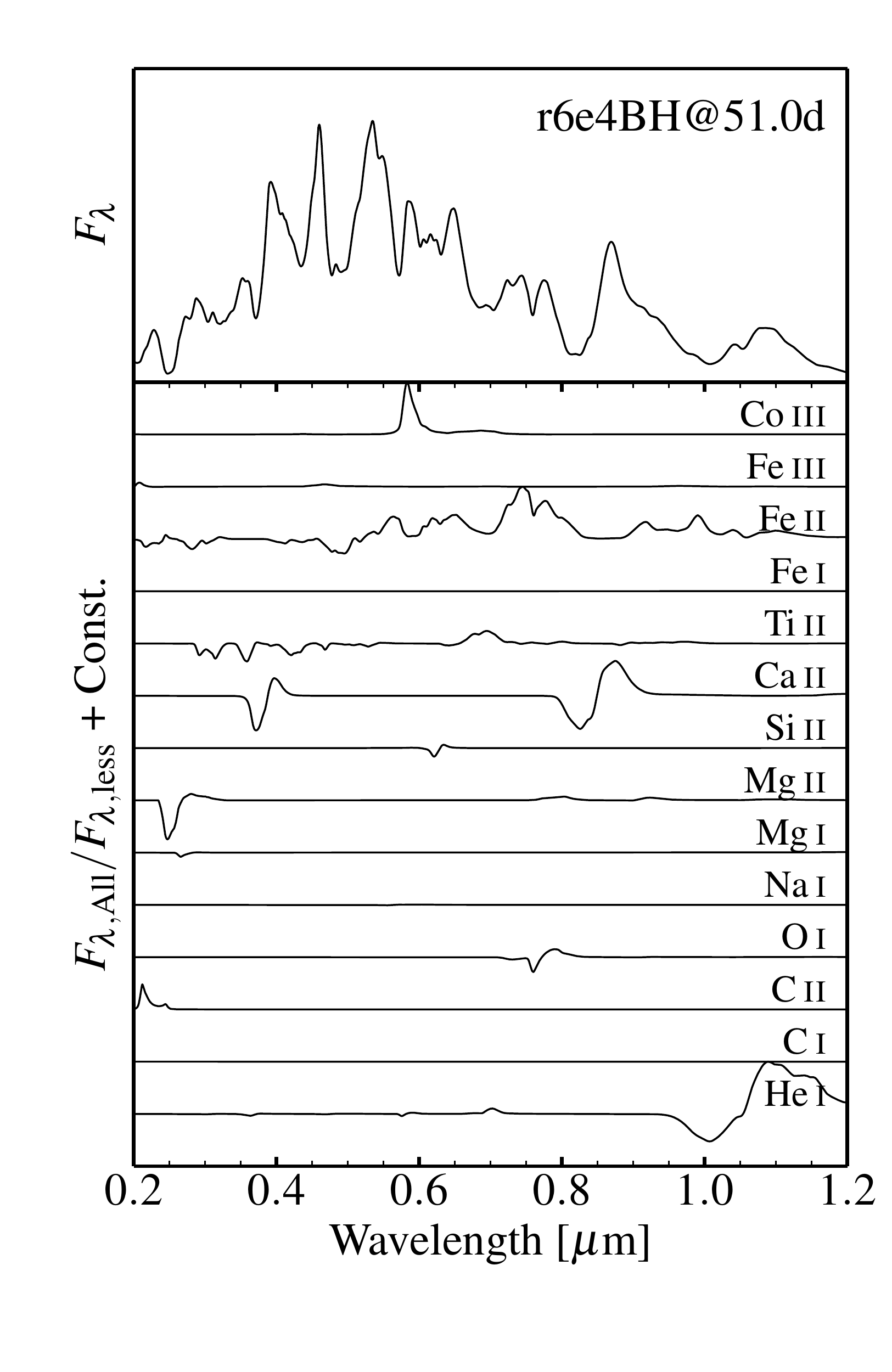,width=7.1cm}
\epsfig{file=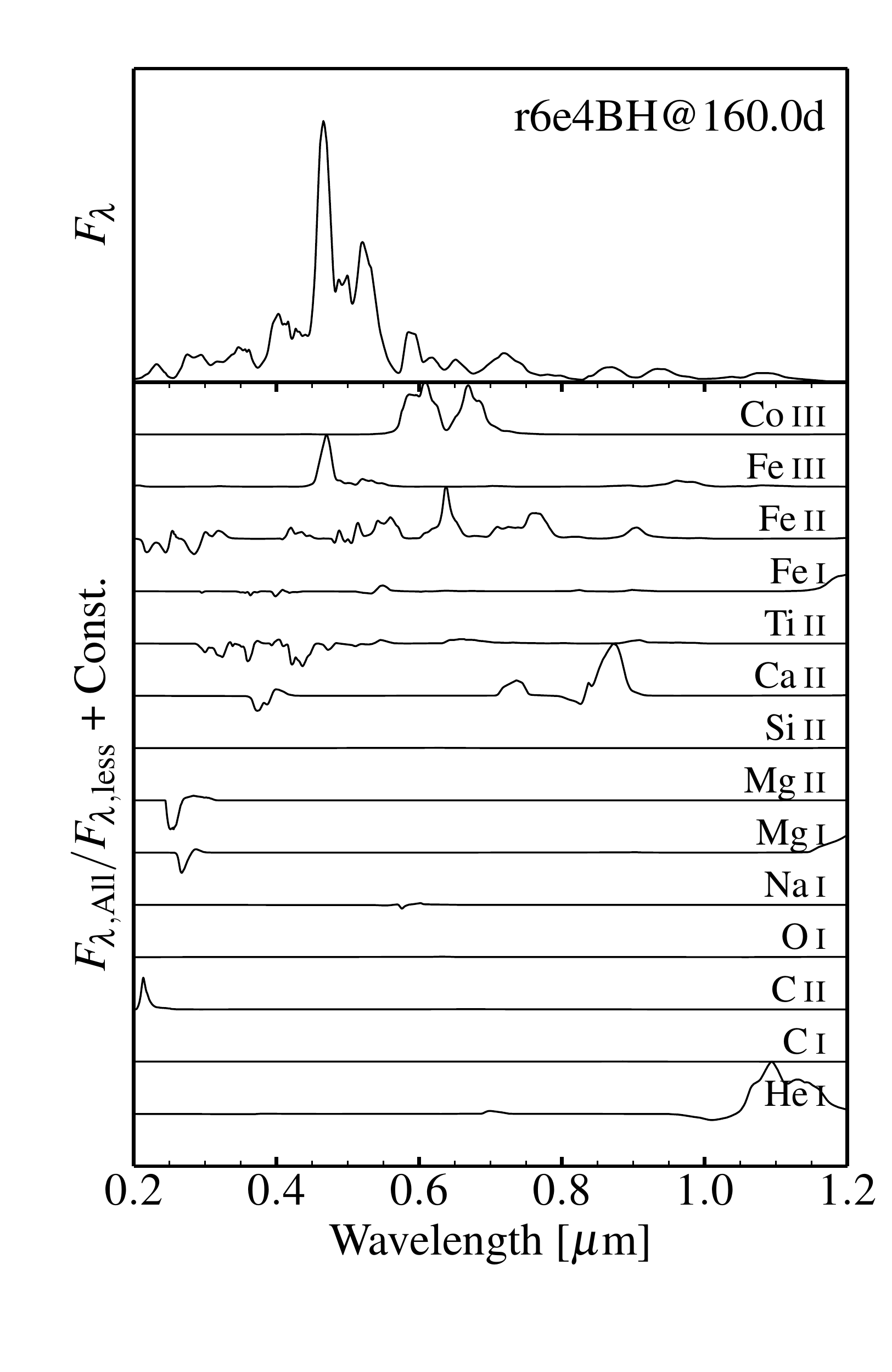,width=7.1cm}
\caption{Same as Fig.~\ref{fig_r6e4bh_ladder_1}, but now for post-explosion times
51.0 (left) and 160.0\,d (right).
\label{fig_r6e4bh_ladder_2}
}
\end{figure*}

\label{lastpage}

\end{document}

%% file: tab_prog_prop.tex
\begin{table*}
\caption{Properties of the models evolved with \mesa, some corresponding
to the initial conditions on the ZAMS and others to the onset of core collapse.
All models start from a 40\,\msun\ star on the ZAMS, but with different initial
rotation rates and metallicities.
\label{tab_prog_prop}}
\begin{center}
\begin{tabular}{
l@{\hspace{2mm}}c@{\hspace{2mm}}c@{\hspace{2mm}}
c@{\hspace{2mm}}c@{\hspace{2mm}}c@{\hspace{2mm}}
c@{\hspace{2mm}}c@{\hspace{2mm}}c@{\hspace{2mm}}
c@{\hspace{2mm}}c@{\hspace{2mm}}c@{\hspace{2mm}}
c@{\hspace{2mm}}c@{\hspace{2mm}}c@{\hspace{2mm}}}
\hline
 Model &       $Z_{\rm init}$ &          $\Omega/\Omega_{\rm crit}$ &                   Age &         $T_{\rm eff}$ &             $L_\star$ &             $R_\star$ &       $M_{\rm final}$ &      $M_{\rm Si,c}$ &      $M_{\rm Fe, c}$ &    $X_{\rm He,s}$ &     $X_{\rm C,s}$ &     $X_{\rm N,s}$ &     $X_{\rm O,s}$ &    $X_{\rm Si,s}$ \\
  &  &  & [Myr]   & [kK] & [\lsun] & [\rsun]    & [\msun] & [\msun] & [\msun] &  & & & & \\
\hline
r0 &               0.0162 &                0 &           5.06 &              166 &              3.39(5) &              0.70 &                11.4 &              1.59 &              1.57 &              1.88(-1) &              5.14(-1) &              0.0 &              2.79(-1) &              5.03(-4) \\
r4 &               0.0162 &               0.4 &             5.44 &              200 &              3.28(5) &              0.48 &              10.0 &   2.01 &              1.82 &              1.75(-1) &              5.20(-1) &              0.0 &              2.86(-1) &              5.03(-4) \\
r6 &               0.0162 &               0.6 &             6.25 &              146&              1.79(5) &              0.67 &               6.6 &        1.67 &              1.55 &              2.91(-1) &              5.38(-1) &              0.0 &              1.52(-1) &              5.02(-4) \\
r6z &              0.002 &                 0.6 &            7.37 &              145 &              2.88(5) &              0.86 &               9.7 &        2.15 &              1.87 &              9.98(-1) &              8.57(-5) &              1.55(-3) &              2.41(-5) &              3.76(-5) \\
\hline
\end{tabular}
\end{center}
\end{table*}

%% file: tab_sn_prop.tex
\begin{table*}
\caption{Summary of ejecta properties used as initial conditions for the \cmfgen\ calculations.
All models derive from pre-SN \mesa\ models named r0, r4, r6, or r6z (see Table~\ref{tab_prog_prop}).
Model r6e4BH forms a 4.04\,\msun\ BH -- all other models leave behind a 1.6--1.9\,\msun\ neutron
star.
\label{tab_sn_prop}}
\begin{center}
\begin{tabular}{l@{\hspace{4mm}}
c@{\hspace{4mm}}
c@{\hspace{4mm}}c@{\hspace{4mm}}c@{\hspace{4mm}}
c@{\hspace{4mm}}c@{\hspace{4mm}}c@{\hspace{4mm}}
c@{\hspace{4mm}}c@{\hspace{4mm}}c@{\hspace{4mm}}
}
\hline
       Model &  $M_{\rm r}$ &    $M_{\rm e}$ &      $E_{\rm kin}$ &       $<V_{\rm m}>$ &      He &    C &      O &    Si &  Ca &  \iso{56}Ni \\
               & [\msun]  &  [\msun]      &             [\foe]              &         [\kms]            &     [\msun]      &[\msun]      &[\msun]      &[\msun]      &[\msun]      &[\msun]      \\
\hline
%    r0e1 &   1.88  &    9.52 &         1.14 &         3460 &         0.178 &         1.328 &         5.388 &         0.041 &        0.0016 &         0.085 \\
    r0e2 &   1.71  &    9.69 &         4.12 &         6530 &         0.181 &         1.326 &         5.471 &         0.112 &        0.0061 &         0.122 \\
%  r0e2m &   1.70  &    9.70 &         4.12 &         6530 &         0.193 &         1.322 &         5.453 &         0.102 &        0.0067 &         0.129 \\
%    r0e3 &   1.62  &    9.78 &         8.20 &         9184 &         0.182 &         1.303 &         5.553 &         0.148 &        0.0081 &         0.140 \\
    r0e4 &   1.54  &    9.86 &        12.31 &        11210 &         0.205 &         1.298 &         5.590 &         0.181 &        0.0098 &         0.172 \\
 \hline
 r4e4 &     1.88   & 8.12 &        13.44 &        12900 &         0.286 &         1.302 &         3.852 &         0.458 &         0.0410 &         0.583 \\
\hline
% r6e1 &    1.93    & 4.66 &         1.17 &         5030 &         0.291 &         1.066 &         2.122 &         0.116 &        0.0035 &         0.097 \\
% r6e2 &    1.71    & 4.88 &         4.28 &         9390 &         0.295 &         1.054 &         2.124 &         0.242 &         0.0172 &         0.196 \\
 r6e4 &    1.62    & 4.97 &        12.41 &        15840 &         0.324 &        1.051 &         2.072 &         0.315 &         0.022 &         0.300 \\
\hline
r6ze4 &   1.99    &  7.70 &        13.70 &        13370 &         1.453 &         0.822 &         3.017 &         0.429 &         0.0386 &         0.696 \\
%r6e4m &   1.62    &  4.97 &        12.41 &        15840 &         0.326 &         1.053 &         2.060 &         0.311 &         0.0221 &         0.300 \\
\hline
r6e4BH&    4.04    & 2.55 &        11.63 &        21420 &         0.456 &         0.910 &         0.515 &         0.051 &        0.0036 &         0.435 \\
\hline
\end{tabular}
\end{center}
\end{table*}